\documentclass[]{spie}  

\usepackage{amsmath,amsfonts,amssymb}
\usepackage{graphicx}
\usepackage{tabularx}
\usepackage[shortlabels]{enumitem}
\usepackage{float}
\usepackage{booktabs}
\usepackage{subcaption}
\usepackage[colorlinks=true, allcolors=blue]{hyperref}
\usepackage{array}
\newcolumntype{Y}{>{\centering\arraybackslash}X}

\title{Performance Testing of a Trillium-based 21-Positioner Module for Stage-5 Telescopes}

\author[a]{Oliver Pineda Suárez}
\author[a]{Jonathan Wei}
\author[a]{Malak Galal}
\author[a]{Melina Daniilidis}
\author[a]{Maxime Rombach}
\author[a]{Sébastien Pernecker}
\author[a]{Jean-Paul Kneib}

\affil[a]{Institute of Physics, Laboratory of Astrophysics,  École Polytechnique Fédérale de Lausanne (EPFL), Observatoire de Sauverny, CH-1290 Versoix, Switzerland}

\authorinfo{Further author information: (Send correspondence to O.P.S. and J.W.)\\O.P.S.: E-mail: oliver.pinedasuarez@epfl.ch\\  J.W.: E-mail: jonathan.wei@epfl.ch}

\begin{document} 
\maketitle

\begin{abstract}
Evaluating the performance metrics of $\theta - \phi$ robotic positioners is crucial to gather insights into the system’s behavior and ensure their reliability during operation in Stage-5 telescopes such as the Chinese MUST, the American Spec-S5 and the European WST. With a careful analysis of these metrics, a comprehensive characterization of the system’s strengths is obtained, alongside a clear identification of aspects requiring improvement. Thus, providing potential avenues for streamlining and optimizing its design and functionality. In this paper we present the results of the positioning performance and angular tilt tests conducted on 6.2-mm-pitch robotic positioner modules developed for high-density fiber positioning in next-generation astronomical systems. The fiber positioners from the tested prototype adopt a mechanical design based on the Trillium open design from the Lawrence Berkeley National Laboratory (LBNL) and were produced by the Japanese company Orbray. The evaluated performance metrics of positioning repeatability, datum repeatability, backlash, non-linearity, and angular tilt were measured and compared to the desired performance for Stage-5 telescopes. The positioners exhibited a generally acceptable performance, although noticeable anomalies affecting several metrics were identified and will require mitigation in subsequent prototypes. Nevertheless, the results indicate promising performance suitable for Stage-5 telescope instrumentation, provided that these issues are successfully resolved.
\end{abstract}

\keywords{Positioning Performance, Angular Tilt, $\theta - \phi$ Architecture, 6.2-mm-Pitch, Robotic Fiber Positioner, Stage-5 Telescopes}

\section{INTRODUCTION} \label{intro}

Dense robotic fiber-positioner systems are a key enabling technology for highly multiplexed spectroscopic surveys. By placing thousands of independent robotic positioners on the focal plane, these systems allow simultaneous acquisition of spectra from many astronomical targets. 

Larger and deeper spectroscopic maps of the Universe aim to extend observations to higher redshifts and allow larger volumes and fainter galaxy populations to be studied \cite{bacon2024wstwidefieldspectroscopic}. At the focal-plane level, these scientific goals translate into demanding engineering constraints.
These future Multi-Object Spectrograph (MOS) instruments require increasingly dense arrays of fiber positioners, often with pitches of only a few millimeters. Each positioner must accurately place a fiber on its assigned target, while
avoiding mechanical interference with neighboring positioners whose patrol regions overlap.  

\begin{figure}[H]
  \makebox[\textwidth][c]{%
    \begin{minipage}{0.7\textwidth}
      \centering
      \captionsetup[subfigure]{justification=centering}
      \begin{subfigure}[b]{0.5\textwidth}
        \centering
        \includegraphics[width=\linewidth]{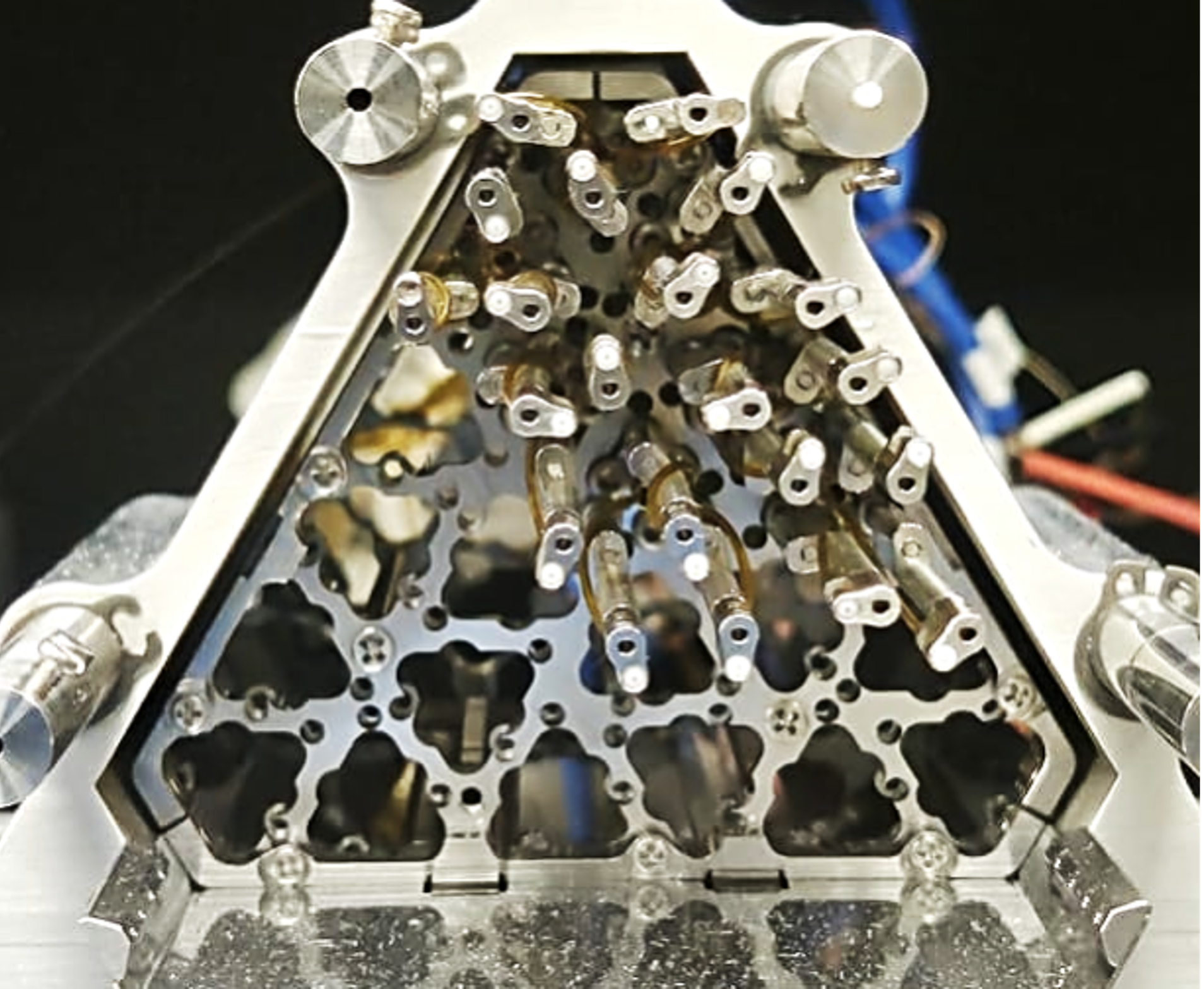}
        \caption{}
        \label{fig:protoorbray}
      \end{subfigure}
      \hspace{.3cm}
      \begin{subfigure}[b]{0.45\textwidth}
        \centering
        \includegraphics[width=\linewidth]{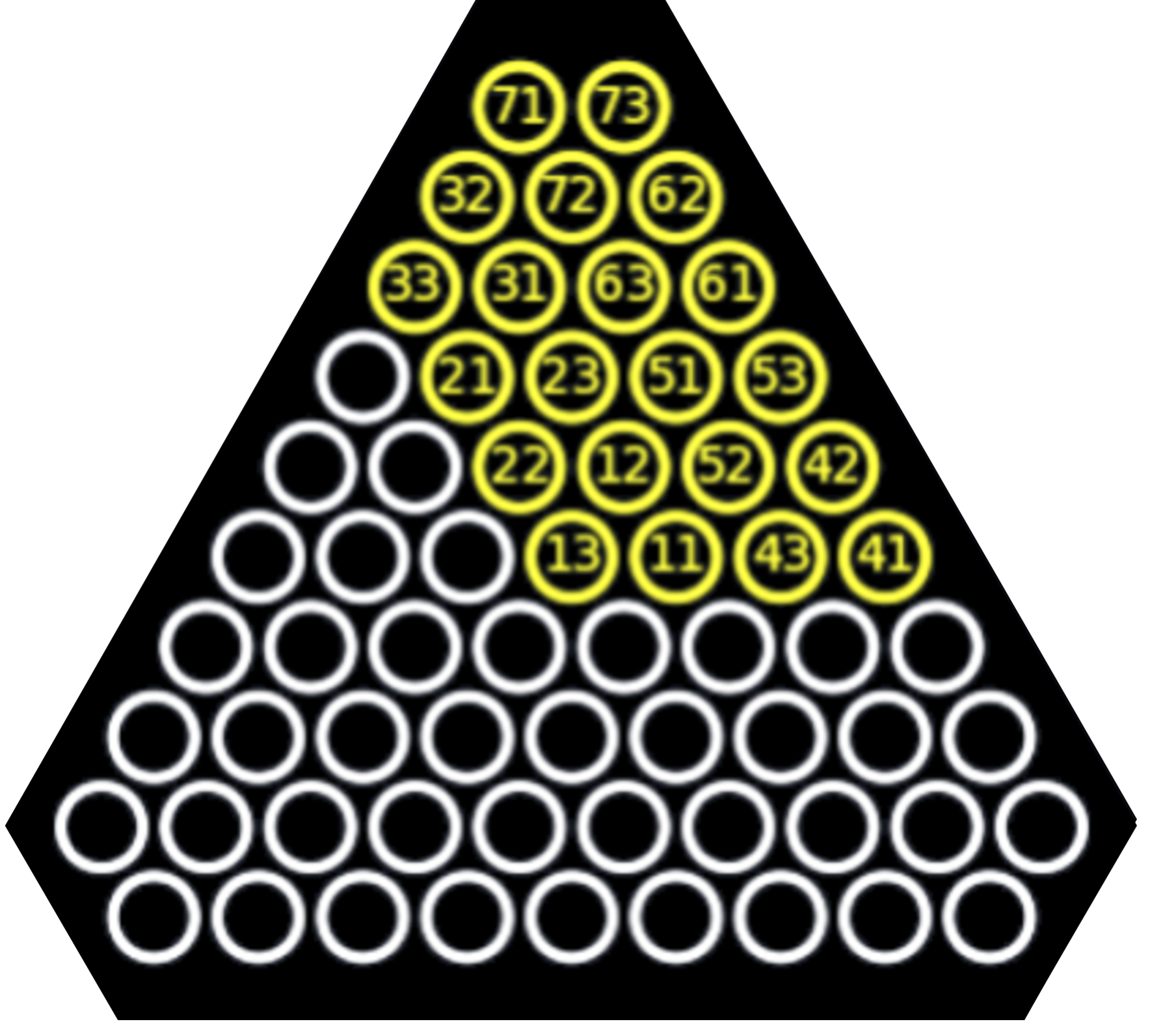}
        \caption{}
        \label{fig:protoorbraynum}
      \end{subfigure}
    \end{minipage}
  }
  \vspace{.1cm}
  \caption{(a) Orbray 21-Fiber Positioner Prototype; (b) Numbering Convention for Fiber-Positioner Identification}
  \label{fig:numconv}
\end{figure}

In this document the performance of the modular fiber positioner system prototype unit shown in Subfigure \ref{fig:protoorbray} will be explored aiming to evaluate its capabilities and identify areas for improvement. These results can be compared to the ones of the module manufactured by MPS \cite{wei2026mps}. This prototype is based on the Trillium design\cite{Silber2022TrilliumII} developed by the team from the Lawerence Berkeley National Laboratory (LBNL) and produced by the Japanese company Orbray Co., Ltd. In this configuration the module is subdivided in "Trilliums" or sets of 3 mechanically jointed positioners with their hardstops shifted by 120 degrees.

Another specific feature of this design is that its axes are coupled so when one rotates the alpha arm, the beta arm will rotate the same number of degrees as the gears are meshed together. This required a software compensation to be implemented in order to rotate the beta arm by the same number of degrees in the opposite direction whenever the alpha arm moves, keeping its relative position to each other.

The prototype from Figure \ref{fig:protoorbray} follow the modular "raft" concept first proposed in Ref. \citenum{SilberModule} adapted to fit 63 robots (using a pitch of 6.2mm) as according to the results obtained in Ref. \citenum{InvestigationsCoverage} this number of positioners is optimal considering coverage and focal surface matching.

To correctly identify the location of each positioner within the module, Subfigure \ref{fig:protoorbraynum} shows the numbering conventions used. These identifiers are used in the subsequent plots and tables throughout this document.

\section{METHODOLOGY}

This section describes the testing methodology used to measure the performance of the fiber-positioner module presented in Section \ref{intro}. The procedures used for these tests are based on the ones described in Ref. \citenum{Galal2025} which were employed for the testing of the previous iteration of these prototypes.

\subsection{XY TEST}

\begin{figure}[H]
    \centering
    \includegraphics[width=0.82\linewidth]{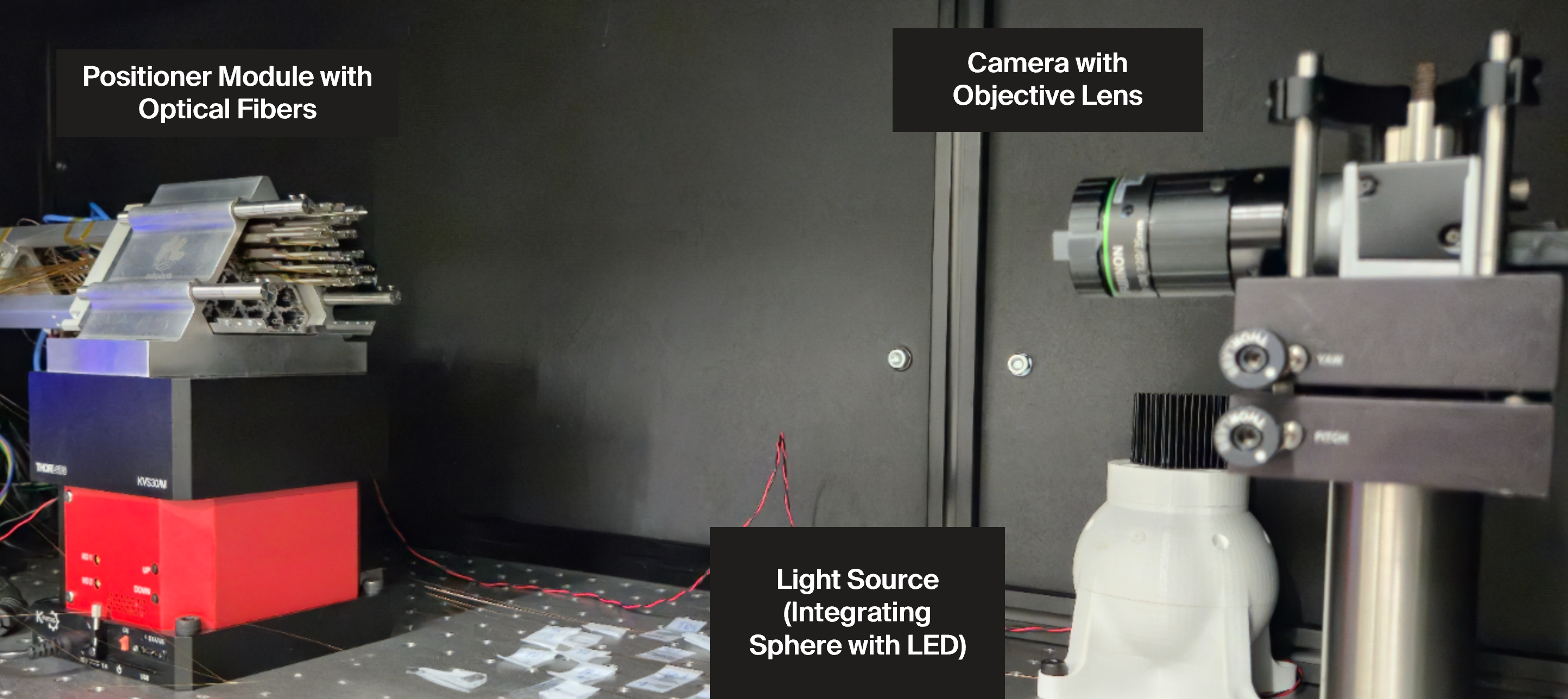}
    \vspace{.1cm}
    \caption{XY Positioning Performance Test Bench}
    \label{fig:setup}
\end{figure}

In order to assess the precision, reliability, and mechanical behavior of the fiber-positioner in the module the following performance metrics must be measured: repeatability, datum performance, backlash, and non-linearity. These metrics provide an overview of the system’s positioning capabilities and the influence of mechanical effects. Figure \ref{fig:setup} presents the experimental test-bench employed. The methodology used implemented in Ref. \citenum{Galal2025} is based on the one established in Ref. \citenum{kronig_precision_2020}.

Repeatability refers to the system’s ability to consistently return to the same position when repeatedly commanded to move to a specific target while approaching from the same direction (clockwise or counterclockwise). It is evaluated through multiple positioning cycles to a given target and quantified by the deviations from the expected position. Low repeatability errors indicate high precision and provide confidence that the system can be reliably calibrated to meet positioning requirements.

Datum repeatability characterizes the system’s ability to consistently return to the hard-stop reference position used for homing procedures. It is measured by repeatedly driving the system into the datum until the hard stop is reached and recording the resulting fiber tip position. Because this reference underpins homing and calibration, high repeatability is essential for robust and consistent system initialization.

Backlash describes the mechanical play arising from clearances in components such as gears and couplings. It is quantified by measuring the positional shift observed when approaching the same target from opposite directions (clockwise and counterclockwise). Lower backlash corresponds to more predictable and accurate positioning behavior, making its characterization an important component of system performance evaluation.

Non-linearity assesses how closely the actual fiber tip motion follows the expected linear response to commanded positions across the actuator range. It is evaluated by comparing measured centroid positions with expected locations at multiple target points. Deviations from ideal linear behavior, primarily caused by transmission effects and mechanical tolerances, indicate modeling and calibration limitations. Lower non-linearity implies a more predictable system response and reduced residual positioning error.

The data presented in Section \ref{res} represents the mean root-mean-square (RMS) values obtained from five repetitions of the test program. The XY coordinates from the positioners in the module are determined by simultaneously acquiring the illuminated spot from the backlight fiber of the positioner using a camera (Basler acA3800-14µm with a Fujinon HF3520-12M objective lens of a 35 mm focal length) with an exposure time of 50 µs. The centroids of these spots are then calculated using two-dimensional Gaussian fitting. 

\subsubsection{TEST PARALLELIZATION}

\begin{figure}[H]
  \makebox[\textwidth][c]{%
    \begin{minipage}{0.8\textwidth}
      \centering
      \captionsetup[subfigure]{justification=centering}
      \begin{subfigure}[b]{0.32\textwidth}
        \centering
        \includegraphics[width=\linewidth]{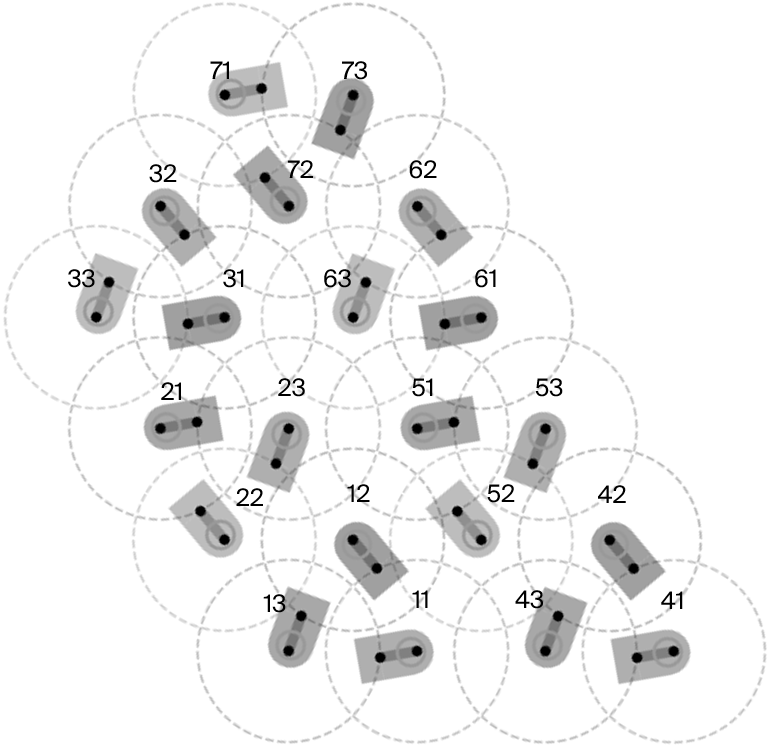}
        \caption{}
        \label{fig:trill_UI}
      \end{subfigure}%
      \hfill
      \begin{subfigure}[b]{0.31\textwidth}
        \centering
        \includegraphics[width=\linewidth]{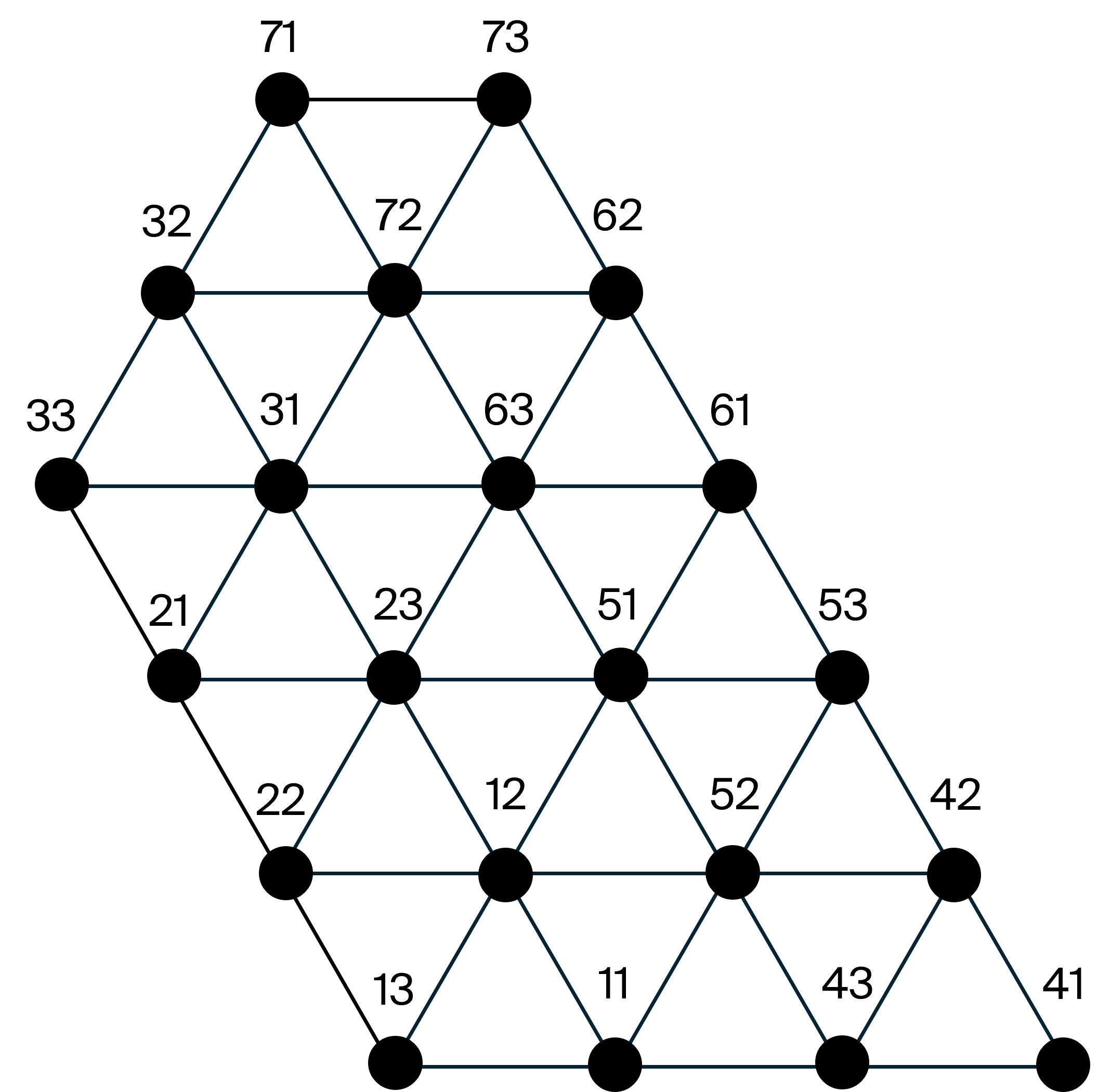}
        \caption{}
        \label{fig:trill_graph}
      \end{subfigure}%
      \hfill
      \begin{subfigure}[b]{0.37\textwidth}
        \centering
        \includegraphics[width=\linewidth]{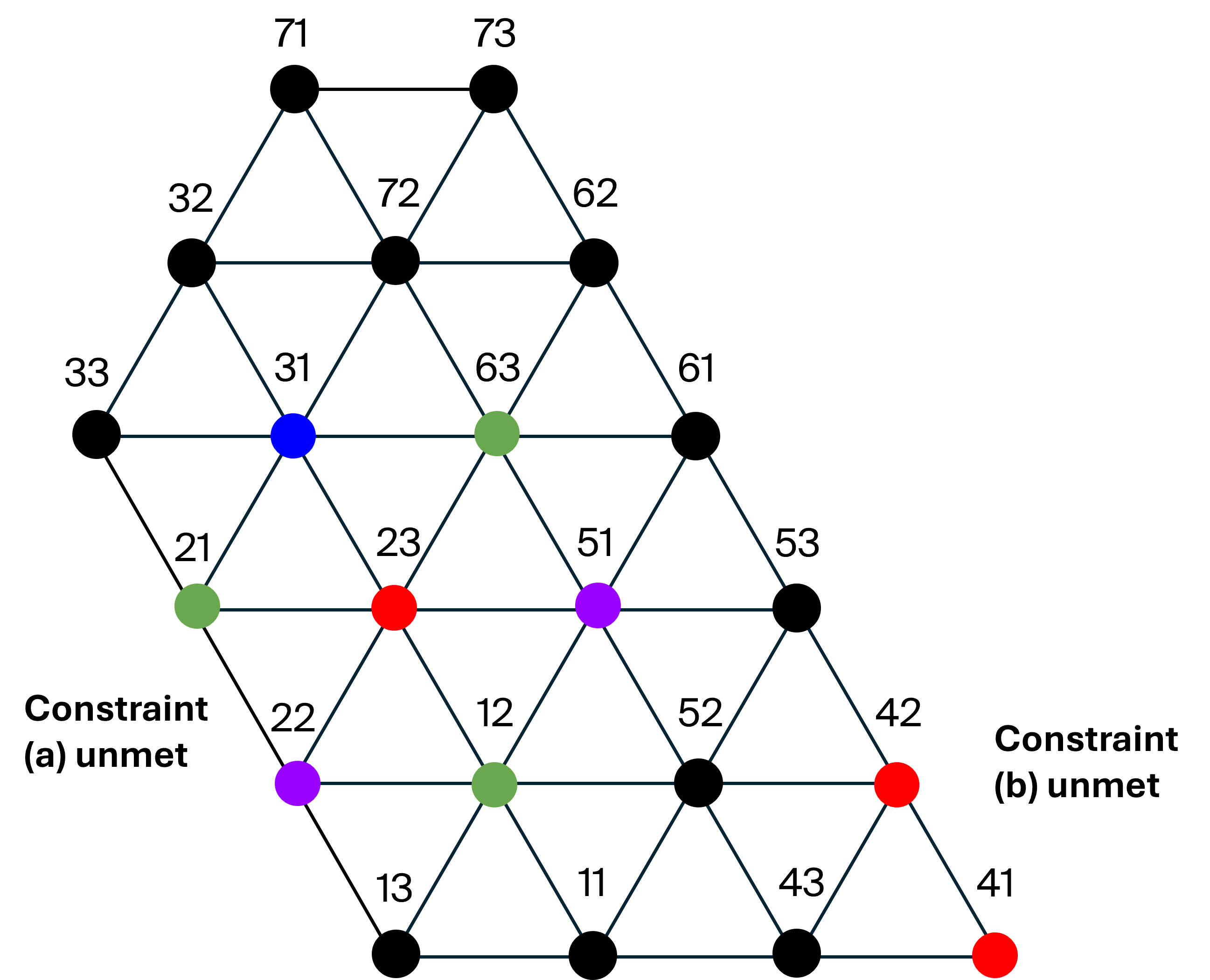}
        \caption{}
        \label{fig:trill_invalid}
      \end{subfigure}
    \end{minipage}
  }
  \vspace{.1cm}
  \caption{(a) Positioner Workspaces (b) Positioner Graph Representation (c) Invalid Grouping Examples}
\end{figure}

Due to the configuration of the Orbray positioners, it is not possible to perform the XY testing simultaneously without causing a collision. Figure \ref{fig:trill_UI} shows the 21 positioners and their workspaces in a dashed line (i.e. the radius is the extension of beta arm by 180 degrees). For a positioner to move freely without collision during testing, there must not be any other positioners within its workspace. From the initial configuration, it is evident that positioners exist within each other's workspaces by default (e.g. Positioners 12 and 53 are in the workspace of Positioner 52). 

One way to mitigate collisions, without increasing the runtime of the XY tests significantly, is to group the positioners in a way such that all positioners in the same group can run their XY tests simultaneously without colliding, and then traverse through each group. To create such groups, the geometry of the Orbray 21-Positioner prototype can be reformulated as a graph coloring problem\cite{Diestel2017}, taken from the field of graph theory. Figure \ref{fig:trill_graph} depicts the graph representation, where each positioner is represented as a node (specifically at the fiber location of the positioner when the beta arm is folded, i.e. the center of each workspace) and edges can be drawn between all of the nodes to connect them. Two nodes are considered adjacent if they share the same edge (e.g. Nodes 51 and 52 are adjacent). A middle node, in this context, is defined as a node that has six adjacent nodes (e.g. Node 52 is a middle mode with adjacent Nodes 51, 53, 12, 42, 11, and 43). 

The nodes can be grouped subject to two constraints. First, no two adjacent nodes can belong to the same group, which prevents positioners with overlapping workspaces from being actuated together. Second, no node can have three adjacent nodes assigned to the same group, which avoids cases where several same-group positioners occupy the patrol region of a single neighboring unit. The second constraint already implies that at least four groups are required for this geometry. 

\begin{figure}[H]
  \makebox[\textwidth][c]{%
    \begin{minipage}{0.95\textwidth}
      \centering
      \captionsetup[subfigure]{justification=centering}
      \begin{subfigure}[b]{0.24\linewidth}
        \centering
        \includegraphics[width=\linewidth]{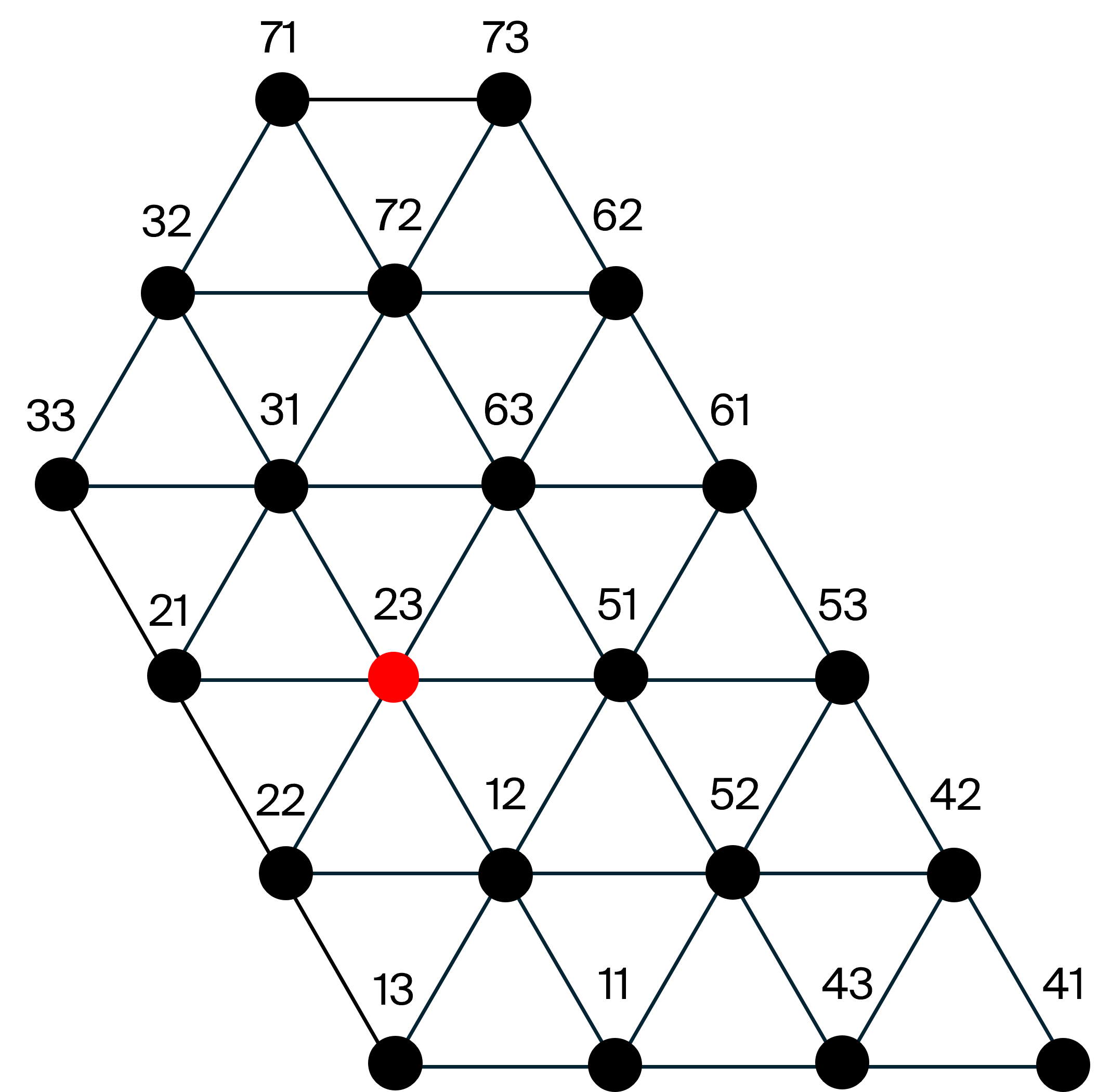}
        \caption{Step 1}
        \label{fig:color_1}
      \end{subfigure}%
      \hfill
      \begin{subfigure}[b]{0.24\linewidth}
        \centering
        \includegraphics[width=\linewidth]{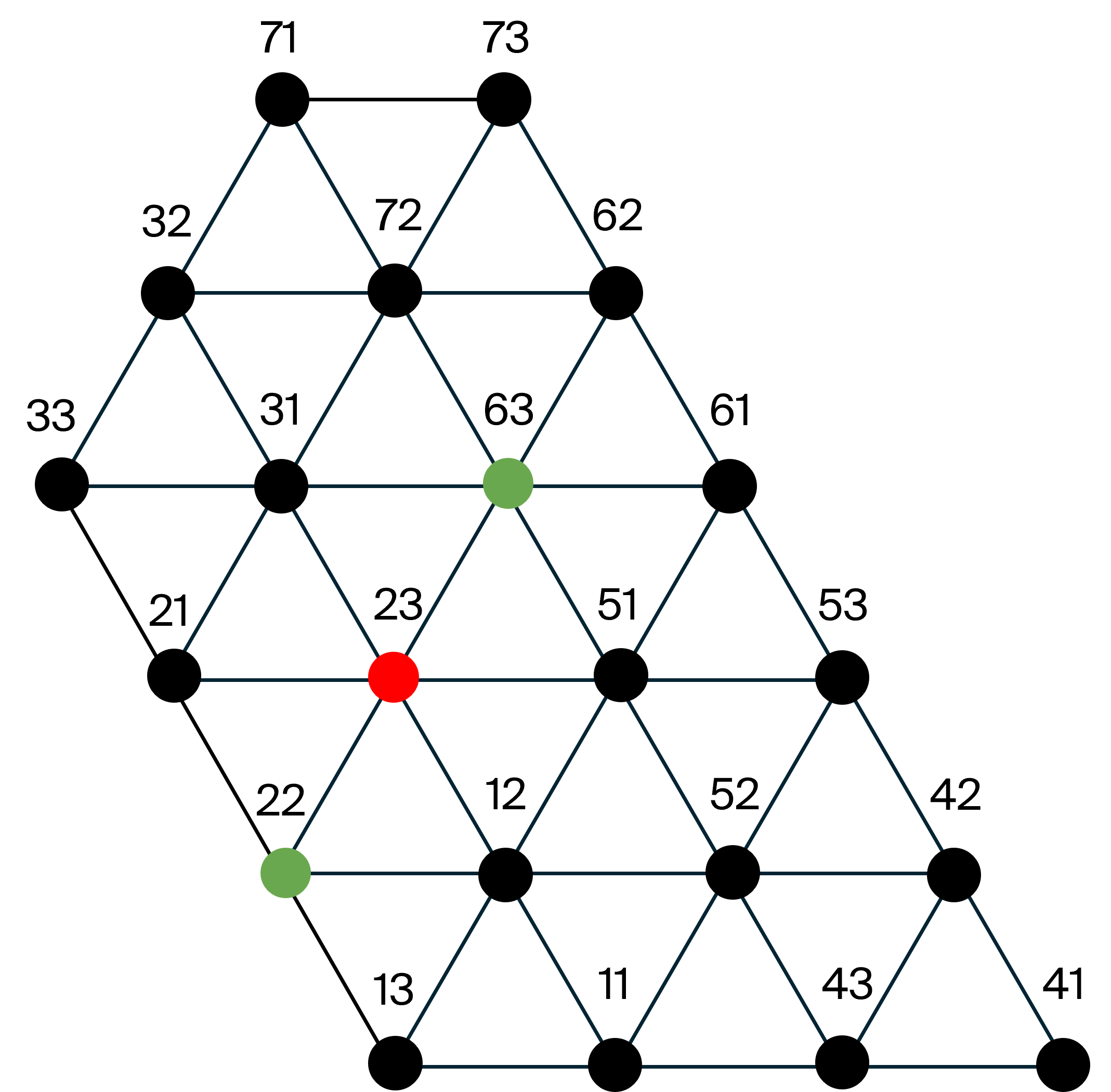}
        \caption{Step 2}
        \label{fig:color_2}
      \end{subfigure}%
      \hfill
      \begin{subfigure}[b]{0.24\linewidth}
        \centering
        \includegraphics[width=\linewidth]{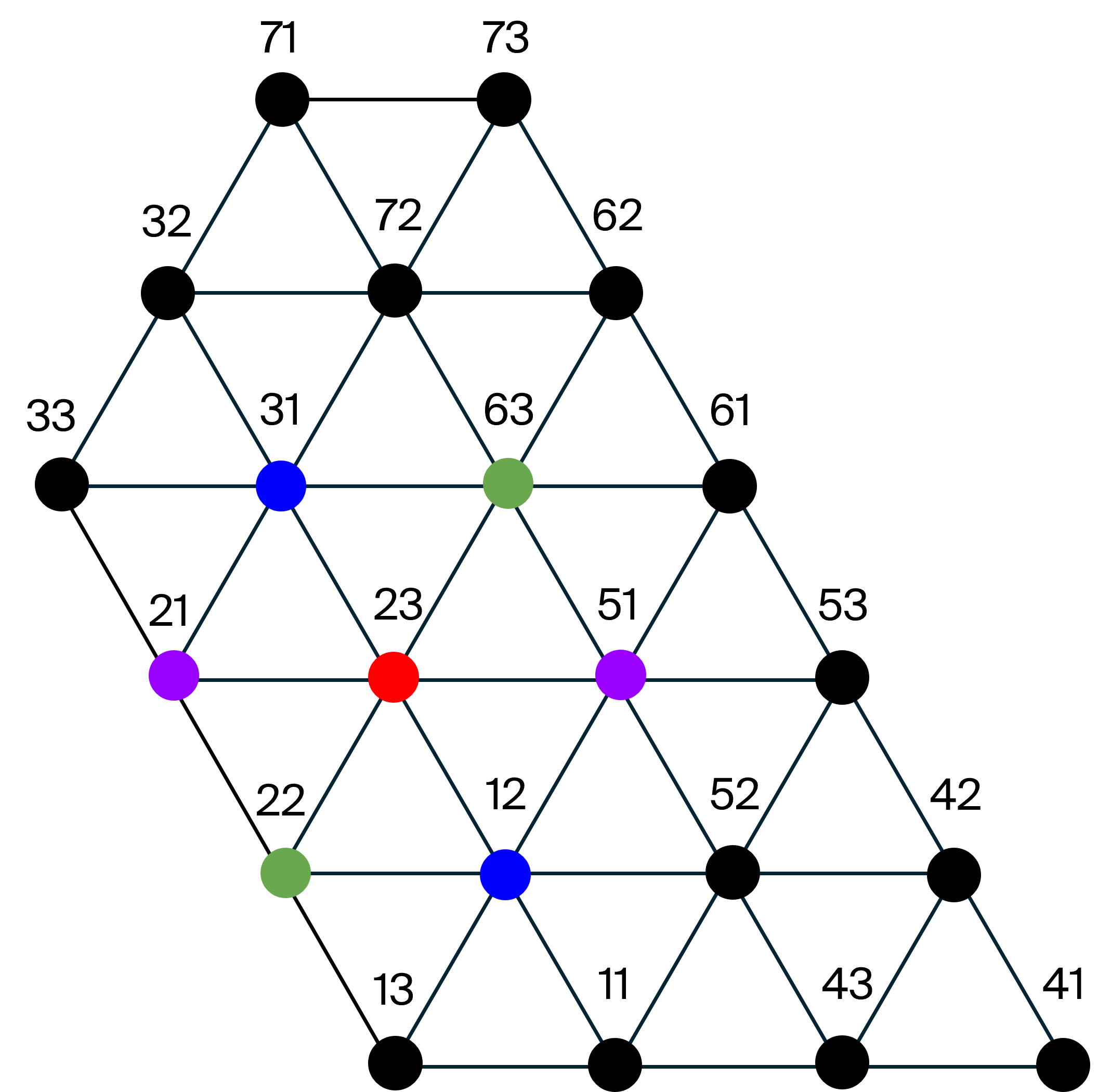}
        \caption{Step 3}
        \label{fig:color_3}
      \end{subfigure}%
      \hfill
      \begin{subfigure}[b]{0.24\linewidth}
        \centering
        \includegraphics[width=\linewidth]{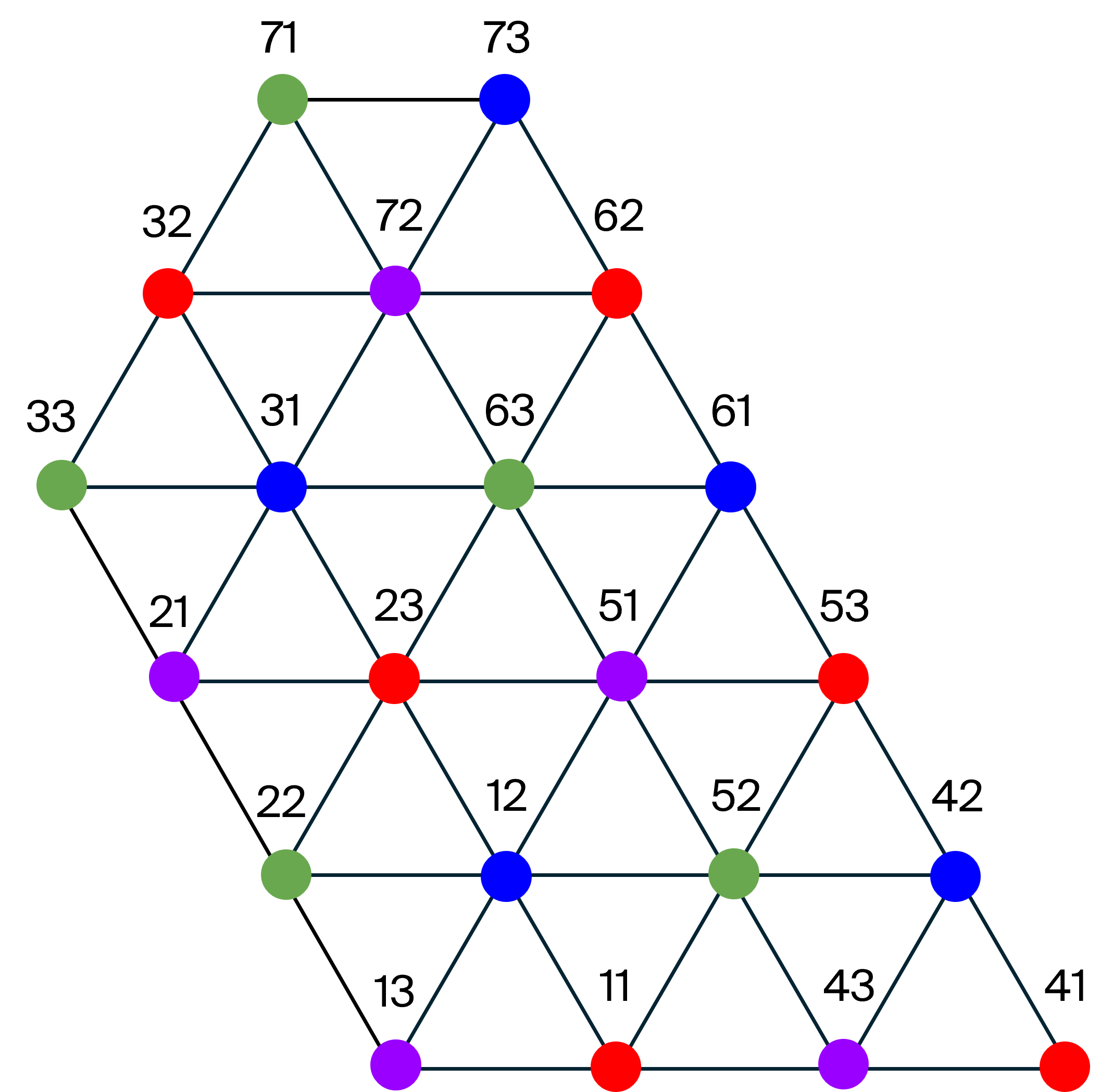}
        \caption{Step 4}
        \label{fig:color_4}
      \end{subfigure}
    \end{minipage}
  }
  \vspace{.1cm}
  \caption{Visualization of the Graph Grouping Steps}
\end{figure}

Figure \ref{fig:trill_invalid} shows examples of invalid groupings. Node 23 violates the second constraint because three of its adjacent nodes, Nodes 63, 21, and 12, are assigned to the same green group. Nodes 42 and 41 violate the first constraint because they are adjacent and both assigned to the red group. To satisfy these constraints, the graph can be grouped following these steps:
\begin{enumerate}
\item Assign a group $r$ to a middle node (Figure \ref{fig:color_1}). 
\item Assign a group $g$ to an adjacent node and the node that is collinearly across from that one, i.e. directly on the other side of the middle node (Figure \ref{fig:color_2}). 
\item Repeat step 2 for groups $b$ and $p$ (Figure \ref{fig:color_3}). 
\item Repeat steps 1-3 until the entire graph has been grouped (Figure \ref{fig:color_4}). At each iteration, the next middle node should have already been grouped from the previous iteration. What remains at this iteration is to group the adjacent nodes of the middle node that have not been grouped yet. 
\end{enumerate}

Geometrically, this means that all positioners from the same group should be at a distance of exactly two times the pitch specified for this modular prototype (6.2mm).

\begin{figure}[H]
  \makebox[\textwidth][c]{%
    \begin{minipage}{0.95\textwidth}
      \centering
      \captionsetup[subfigure]{justification=centering}
      \begin{subfigure}[b]{0.24\linewidth}
        \centering
        \includegraphics[width=\linewidth]{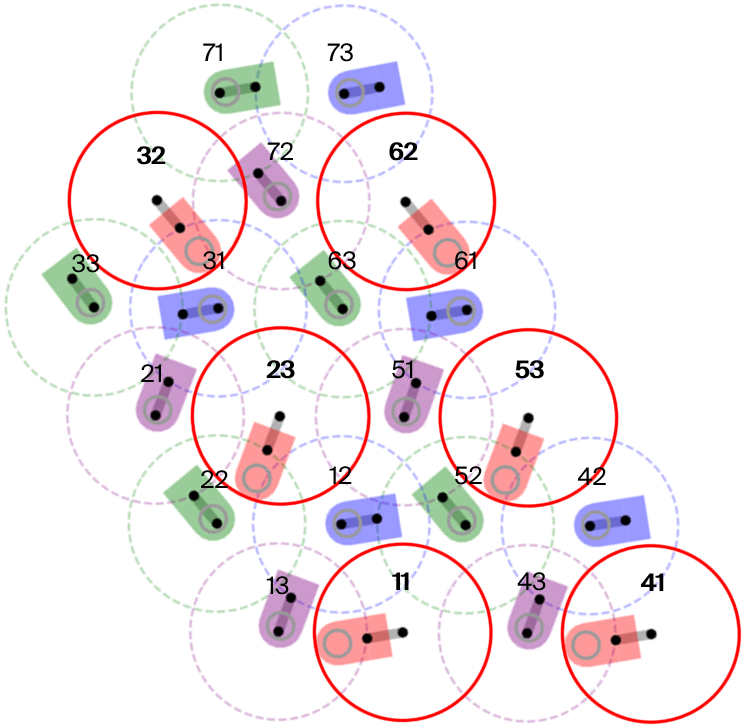}
        \caption{Red Group}
      \end{subfigure}%
      \hfill
      \begin{subfigure}[b]{0.24\linewidth}
        \centering
        \includegraphics[width=\linewidth]{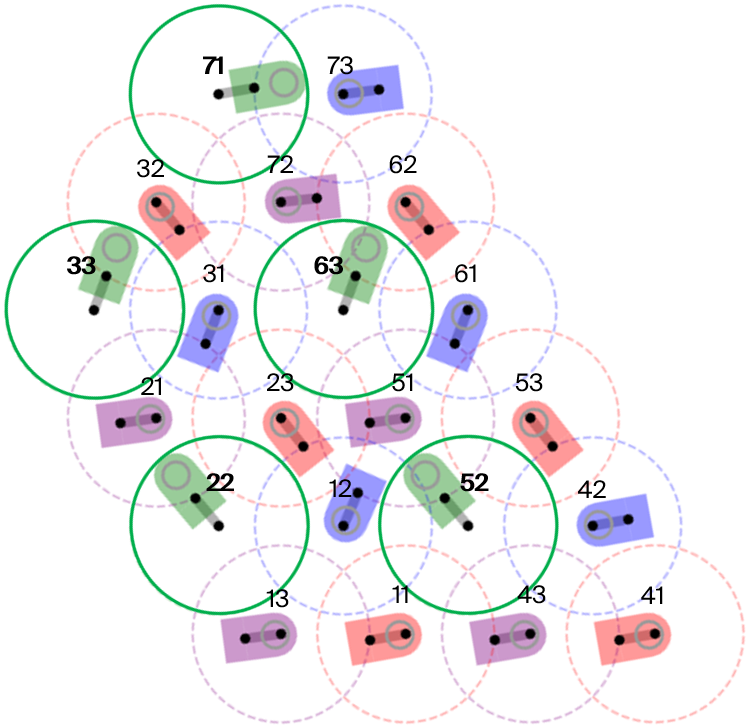}
        \caption{Green Group}
      \end{subfigure}%
      \hfill
      \begin{subfigure}[b]{0.24\linewidth}
        \centering
        \includegraphics[width=\linewidth]{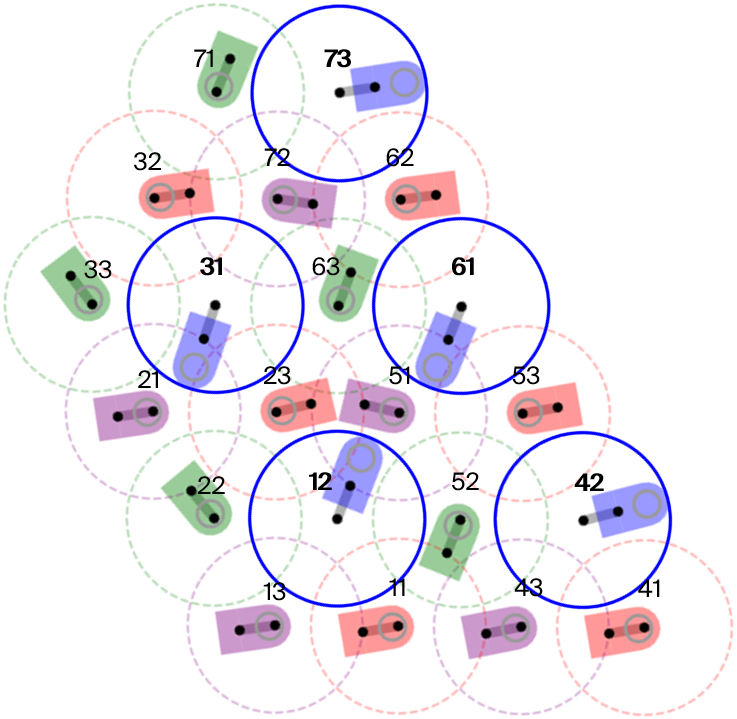}
        \caption{Blue Group}
      \end{subfigure}%
      \hfill
      \begin{subfigure}[b]{0.24\linewidth}
        \centering
        \includegraphics[width=\linewidth]{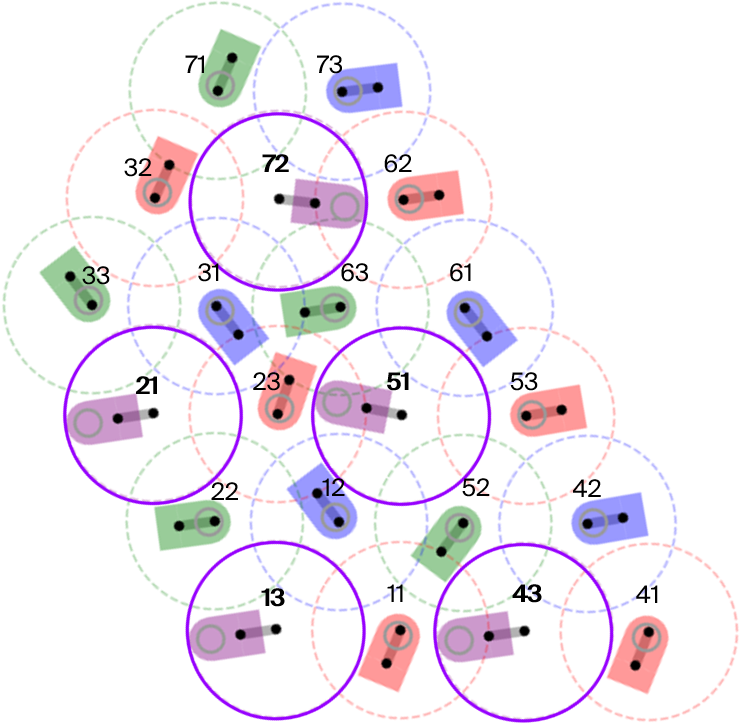}
        \caption{Purple Group}
      \end{subfigure}
    \end{minipage}
  }
  \vspace{.1cm}
  \caption{Positioner Workspaces Isolated by Group}
  \label{fig:color_isolation}
\end{figure}

Once the positioners have been grouped, the near-parallelization occurs from taking the positioners of a certain group, isolating their workspaces (by moving the hindering positioners outside of their workspaces) and performing the XY tests on these isolated positioners. Figure \ref{fig:color_isolation} shows the positioner configurations necessary to isolate the positioner workspaces by group.

\subsection{ANGULAR TILT TEST}

\begin{figure}[H]
  \makebox[\textwidth][c]{\includegraphics[width=0.6\linewidth]{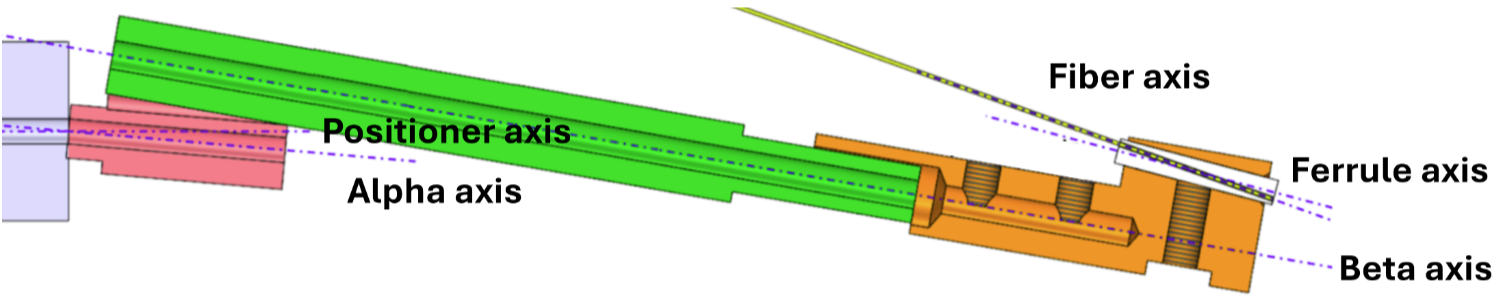}}
  \caption{Angular Tilt Axes in Theta-Phi Fiber-Positioners Schematic from Ref. \citenum{Galal2025}}
  \label{fig:tilt_axes}
\end{figure}

To characterize the alignment, the optical fibers are first mounted on a precision cylinder and rotated to measure the tilt within the ferrule. The tilt between the ferrule axis and the beta arm is obtained by rotating the beta arm, fitting a circle, and extracting its radius. In a similar way, the tilt between the beta and alpha arms is determined by separately rotating each arm, fitting circles to their motions, and computing the angle between the resulting centroids. Finally, the tilt between the alpha arm and the positioner axis is measured using a calibration cylinder as a zero-reference, with its rotation used to fit a circle whose center defines the reference point\cite{Galal2025,10.1117/1.JATIS.6.1.018001}. Figure \ref{fig:tilt_axes} shows a schematic of the mentioned tilt axes.\\

A bi-convex lens (500 mm) with a diffuser screen placed at its focal length lets tilt be measured independently of XY translation. As the fiber source (Thorlabs MINTF4) moves in X and Y in the routine mentioned previously, the beam’s focal point on the diffuser stays fixed. Only of there is tilt present in the fiber or positioner the spot shifts. The tilt-induced displacement is measured as the difference between the shifted and the original points\cite{Galal2025}.

\begin{figure}[H]
    \centering
    \includegraphics[width=0.82\linewidth]{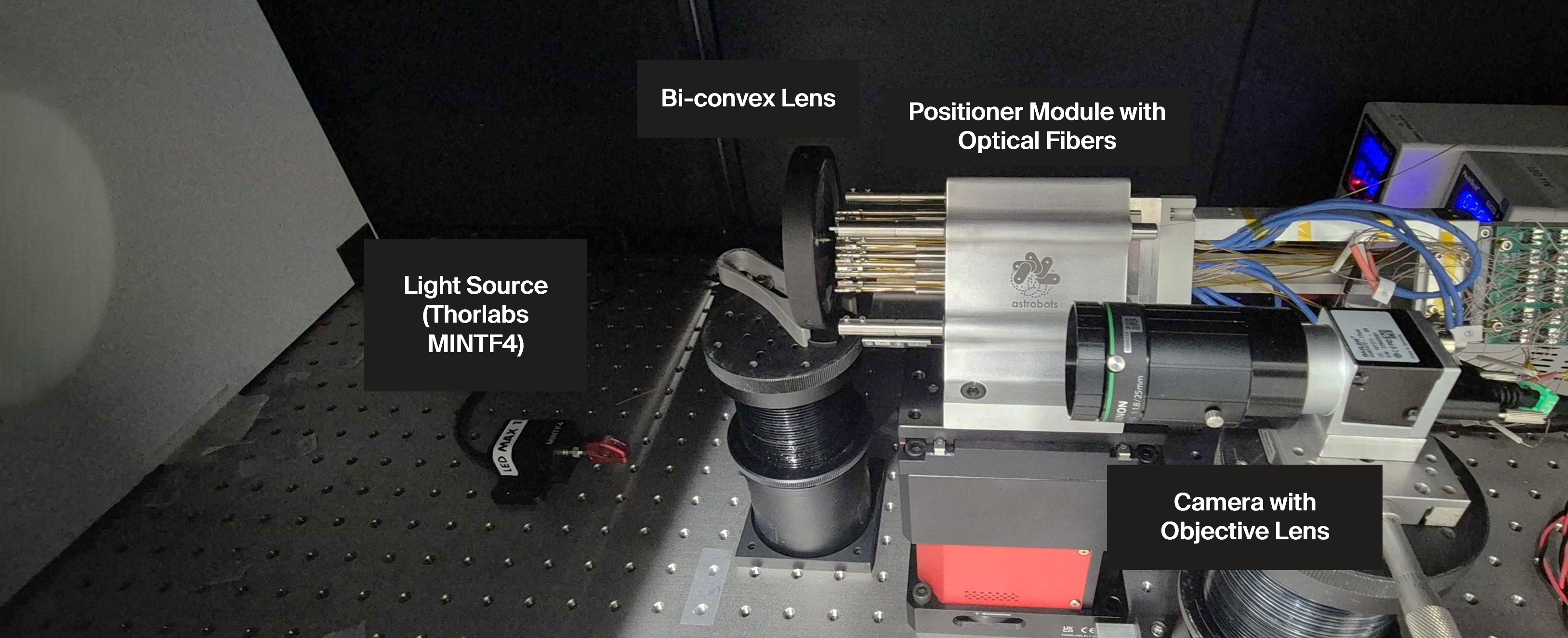}
    \vspace{.1cm}
    \caption{Angular Tilt Test Bench}
    \label{fig:setup2}
\end{figure}

The raw image of the spot (shown in the left of Figure \ref{fig:setup2}) is captured by the camera (Basler acA3800-14µm with a Fujinon CF35ZA-1S objective lens of a 25 mm focal length) and then processed using Gaussian blurring and Otsu’s thresholding to find the spot's centroid. As the alpha and beta arms rotate, the centroids trace a circle, which is fitted using least squares to estimate its center and radius\cite{Galal2025}. Figure \ref{fig:setup2} shows the tilt test bench used to evaluate these positioners.

\section{RESULTS} \label{res}

In this section the performance results obtained from the previously described tests are presented. The results are obtained as the mean from 5 repetitions of the test program.

\subsubsection{REPEATABILITY}

\begin{figure}[H]
  \makebox[\textwidth][c]{%
    \begin{minipage}{1\textwidth}
      \centering
      \captionsetup[subfigure]{justification=centering}
      \begin{subfigure}[b]{0.495\textwidth}
        \centering
        \includegraphics[width=\linewidth]{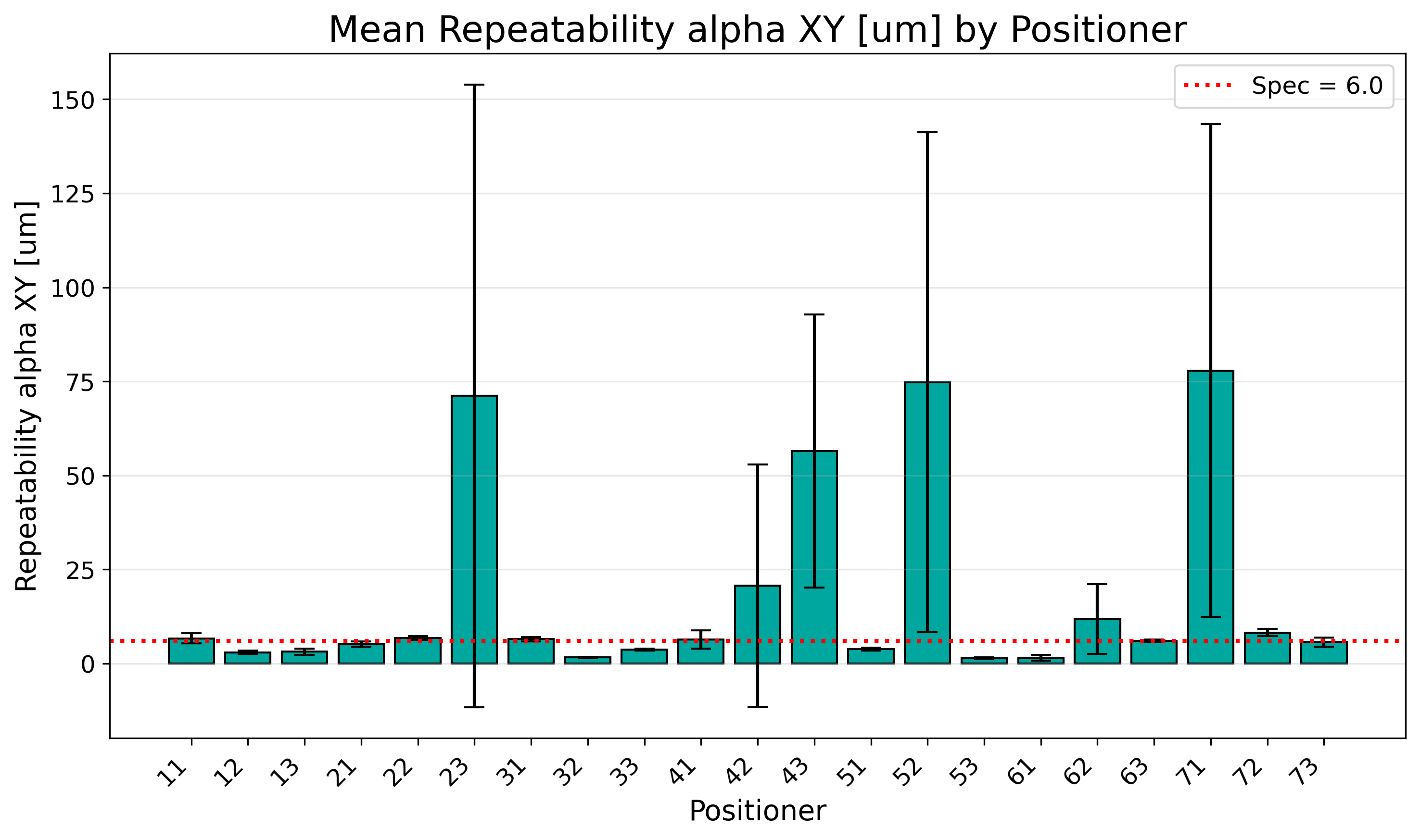}
        \caption{Alpha arm}
        \label{fig:repalpha}
      \end{subfigure}
      \begin{subfigure}[b]{0.495\textwidth}
        \centering
        \includegraphics[width=\linewidth]{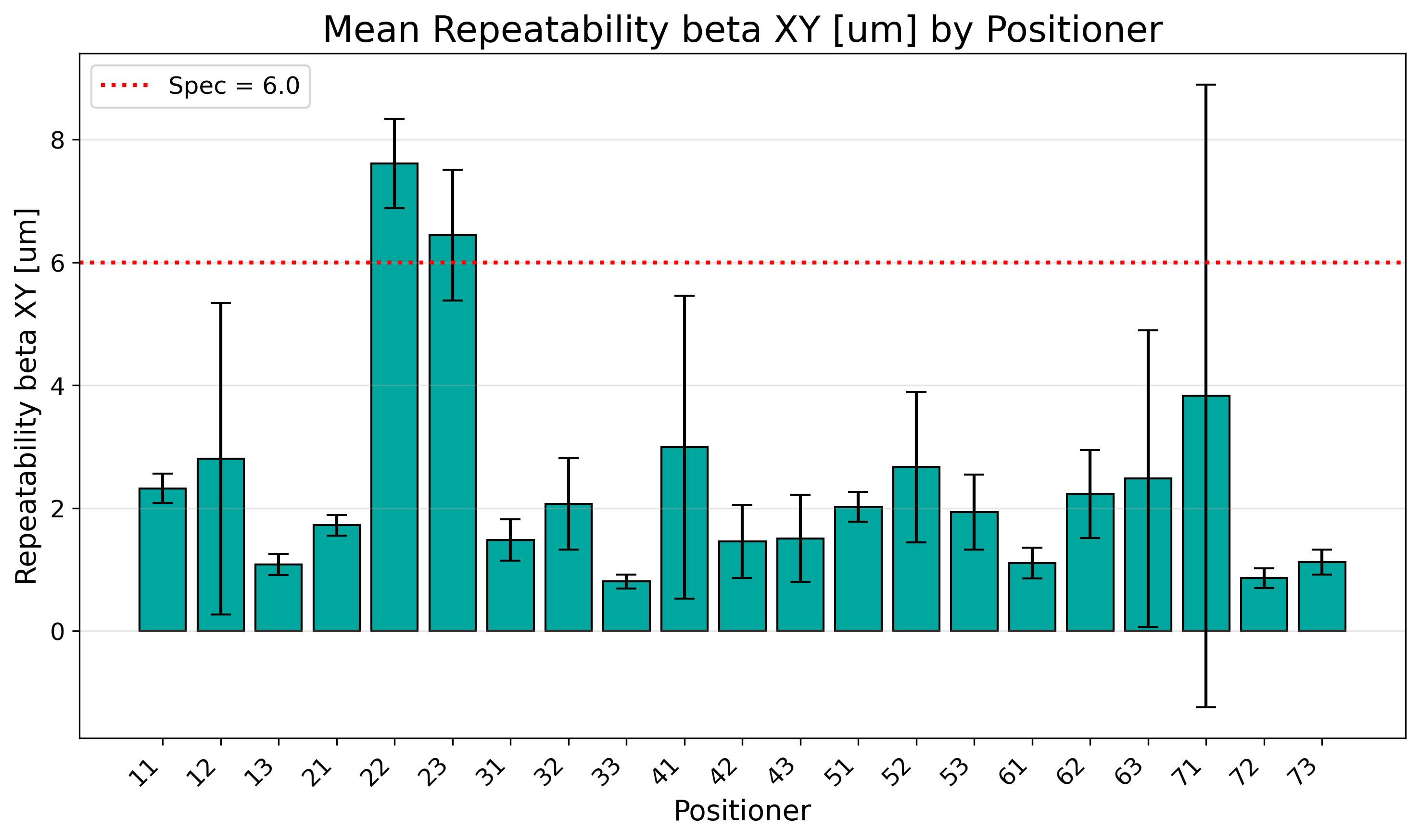}
        \caption{Beta arm}
        \label{fig:repbeta}
      \end{subfigure}
    \end{minipage}
  }
  \vspace{.1cm}
  \caption{Repeatability performance per fiber-positioner. The bar plots show the root-mean-square results. Note: The red dotted line depicts the desired performance \(\le \) 6 µm RMS for each arm.}
  \label{fig:rep}
\end{figure}

To evaluate repeatability, each positioner is subjected to repeated commands driving the alpha and beta arms to a fixed target position in the same direction. The data shown in Figure \ref{fig:rep} correspond to the mean results for both axes.

\subsubsection{DATUM REPEATABILITY}

\begin{figure}[H]
  \makebox[\textwidth][c]{%
    \begin{minipage}{1\textwidth}
      \centering
      \captionsetup[subfigure]{justification=centering}
      \begin{subfigure}[b]{0.495\textwidth}
        \centering
        \includegraphics[width=\linewidth]{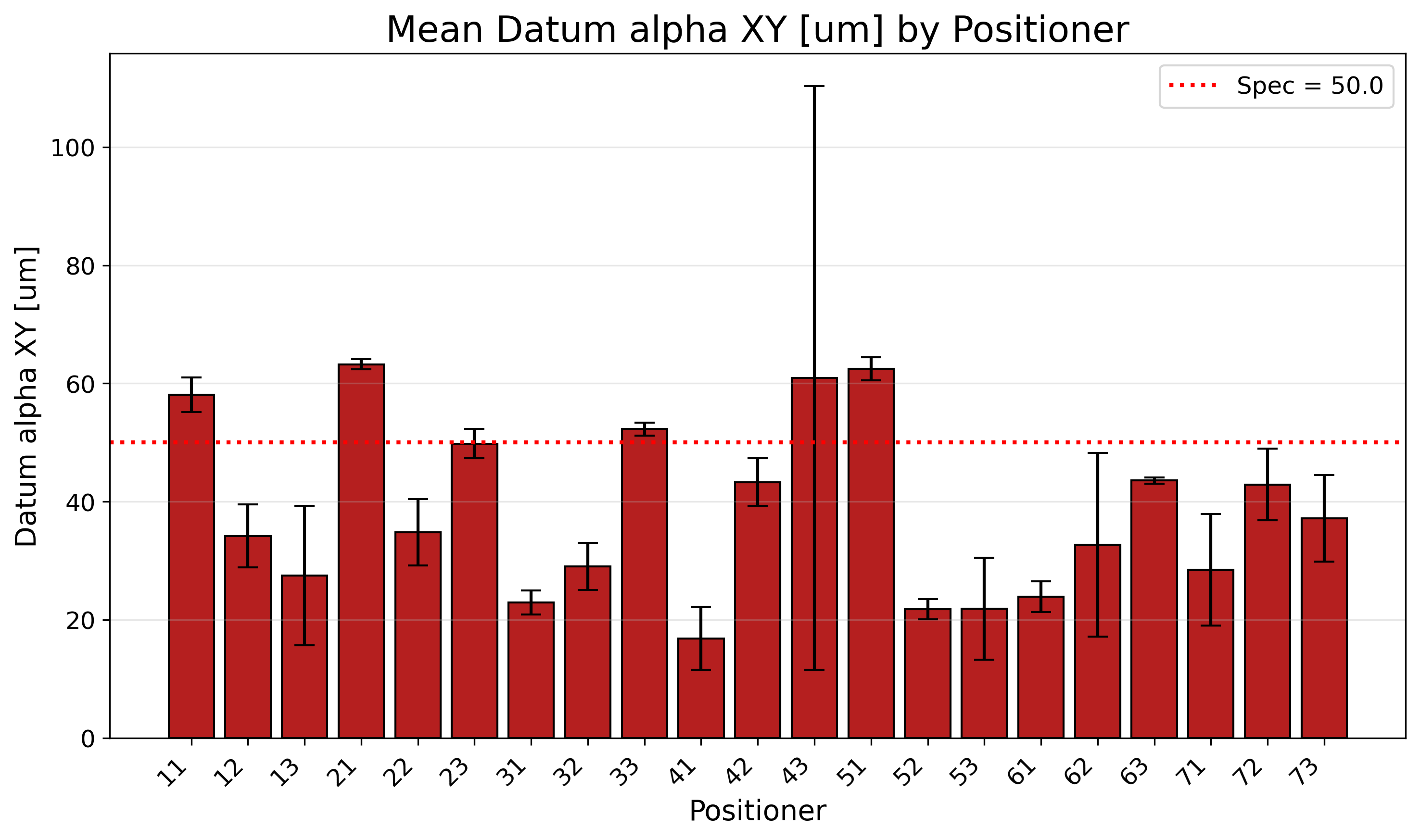}
        \caption{Alpha arm}
        \label{fig:datumalpha}
      \end{subfigure}
      \begin{subfigure}[b]{0.495\textwidth}
        \centering
        \includegraphics[width=\linewidth]{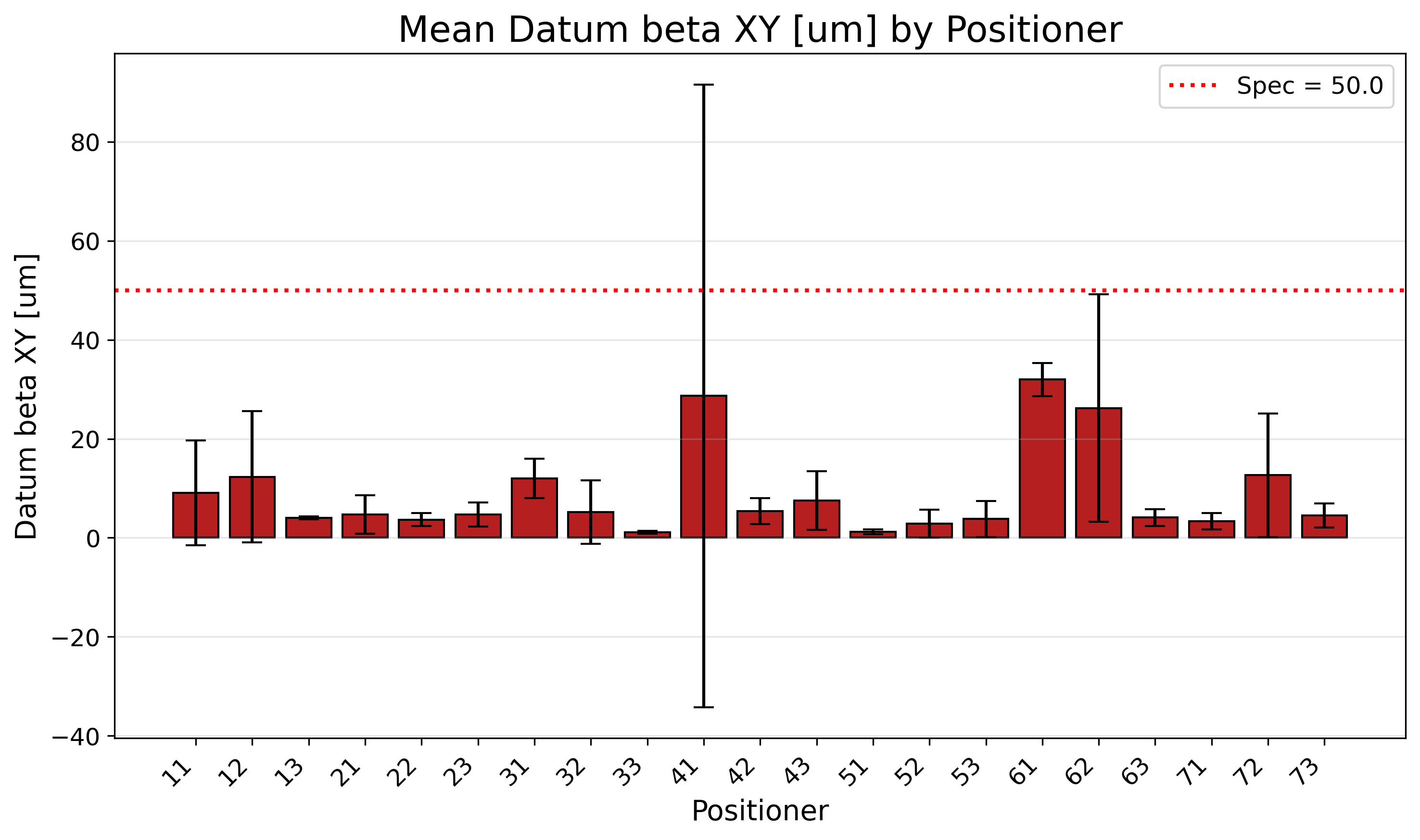}
        \caption{Beta arm}
        \label{fig:datumbeta}
      \end{subfigure}
    \end{minipage}
  }
  \vspace{.1cm}
  \caption{Datum repeatability performance per fiber-positioner. The bar plots show the root-mean-square results. Note: The red dotted line depicts the desired performance  \(\le \) 50 µm RMS for each arm.}
  \label{fig:datum}
\end{figure}

To assess the datum repeatability, the data is acquired by moving the alpha and beta arms to their hard stops. The centroids of the illuminated spots are found using 2D Gaussian fitting, and the root mean square of the repeatability is calculated from 20 iterations. The data in Figure \ref{fig:datum} shows the mean results per each positioner.

\subsubsection{BACKLASH}

\begin{figure}[H]
  \makebox[\textwidth][c]{%
    \begin{minipage}{1\textwidth}
      \centering
      \captionsetup[subfigure]{justification=centering}
      \begin{subfigure}[b]{0.495\textwidth}
        \centering
        \includegraphics[width=\linewidth]{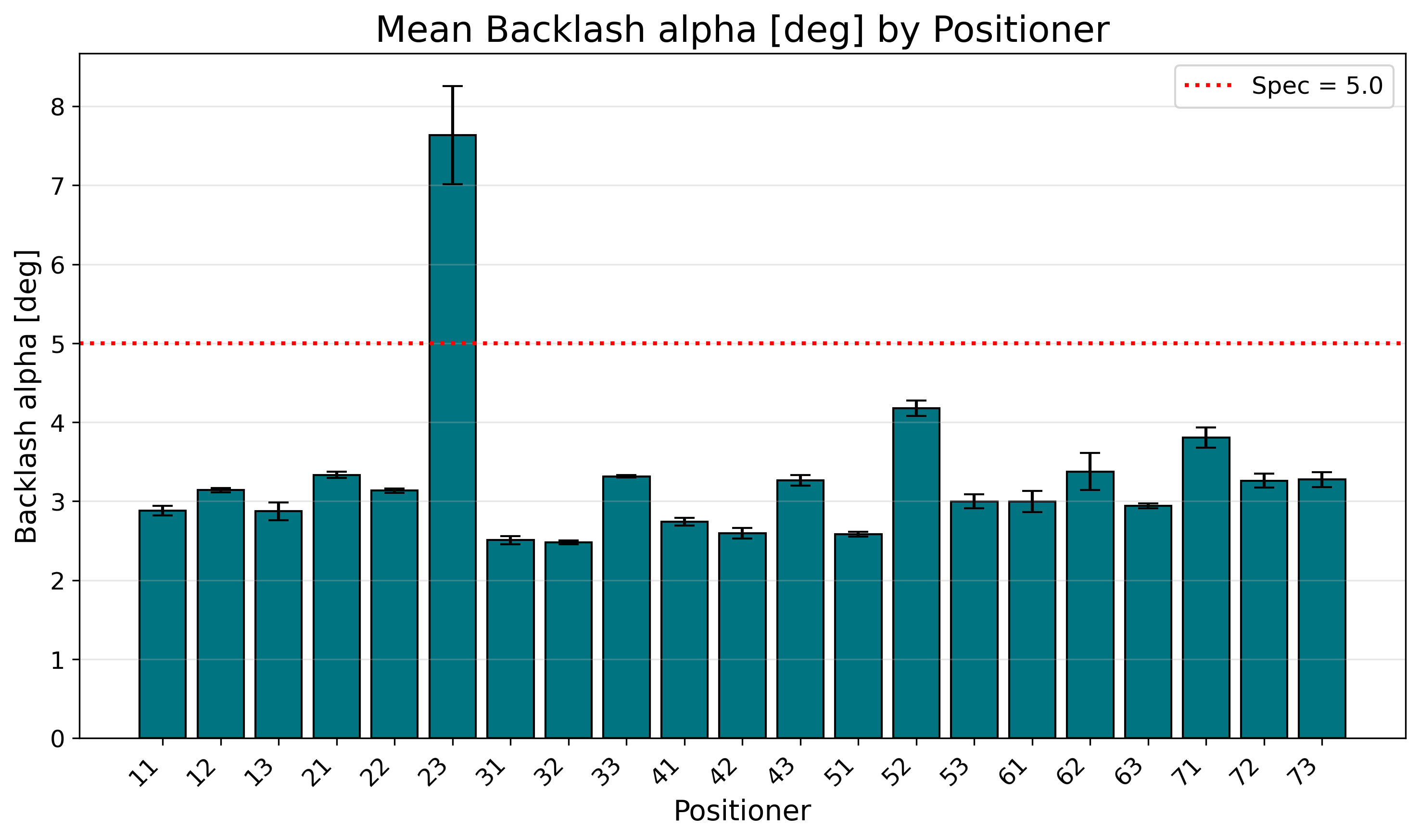}
        \caption{Alpha arm}
        \label{fig:backlashalpha}
      \end{subfigure}
      \begin{subfigure}[b]{0.495\textwidth}
        \centering
        \includegraphics[width=\linewidth]{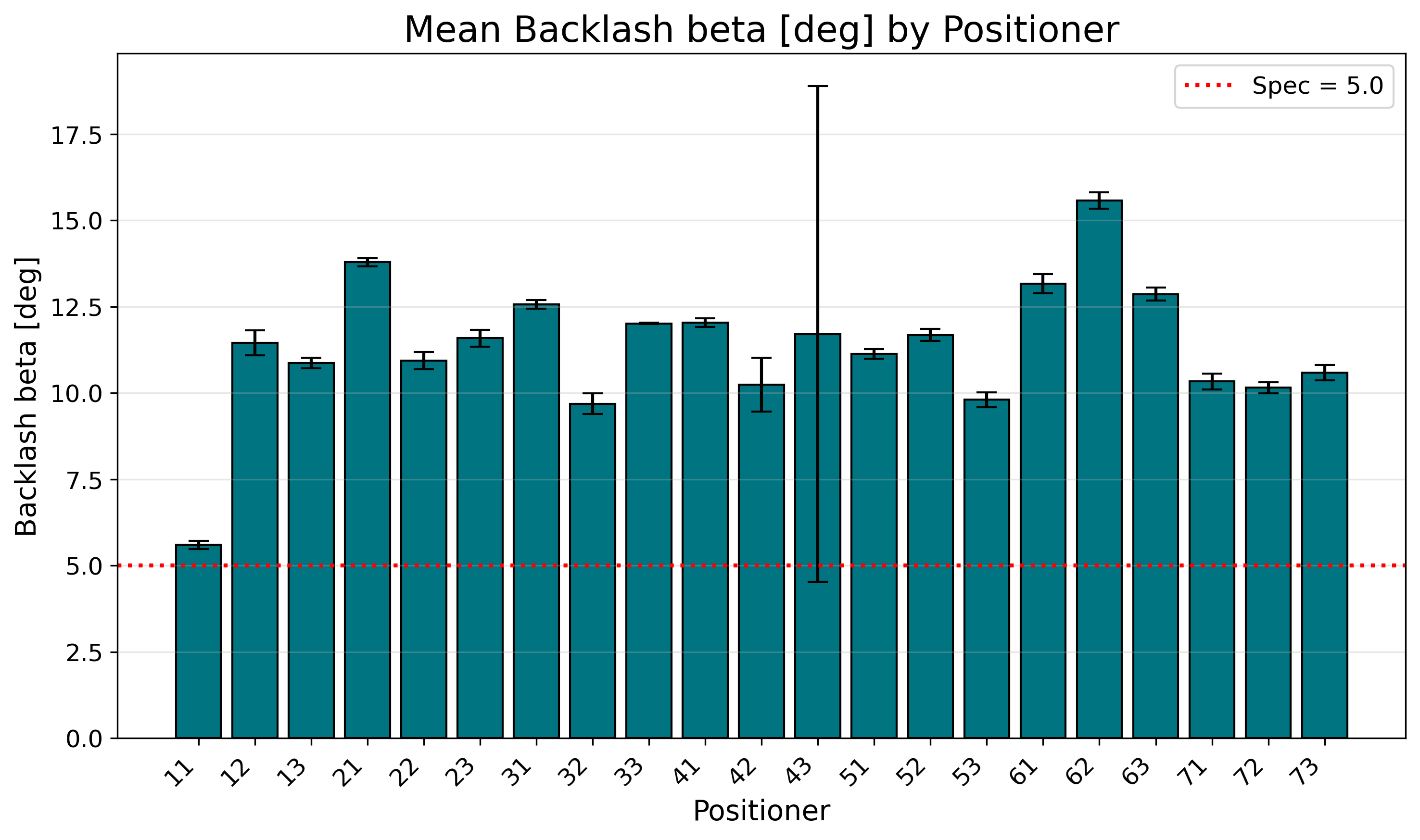}
        \caption{Beta arm}
        \label{fig:backlashbeta}
      \end{subfigure}
    \end{minipage}
  }
  \vspace{.1cm}
  \caption{Calculated backlash per fiber-positioner. The bar plots show the obtained results. Note: The red dotted line depicts the desired performance \(\le \) 5 degrees for each arm.}
  \label{fig:backlash}
\end{figure}

For the backlash measurement, the positioner is commanded to repeatedly move between two positions separated by a reference angular distance for 20 iterations and calculating the centroids of the spots for each position. The backlash is defined as the difference between the measured angular distance of the two centroids and the reference angular distance. Figure \ref{fig:backlash} presents the mean of the results obtained.

\subsubsection{NON LINEARITY}

The non-linearity of each fiber-positioner was tested by performing 1 degree movements along the motion range from each one of its arms from its hard stop in one direction to its hard stop in the opposite direction.

\begin{figure}[H]
    \centering
    \includegraphics[width=1\linewidth]{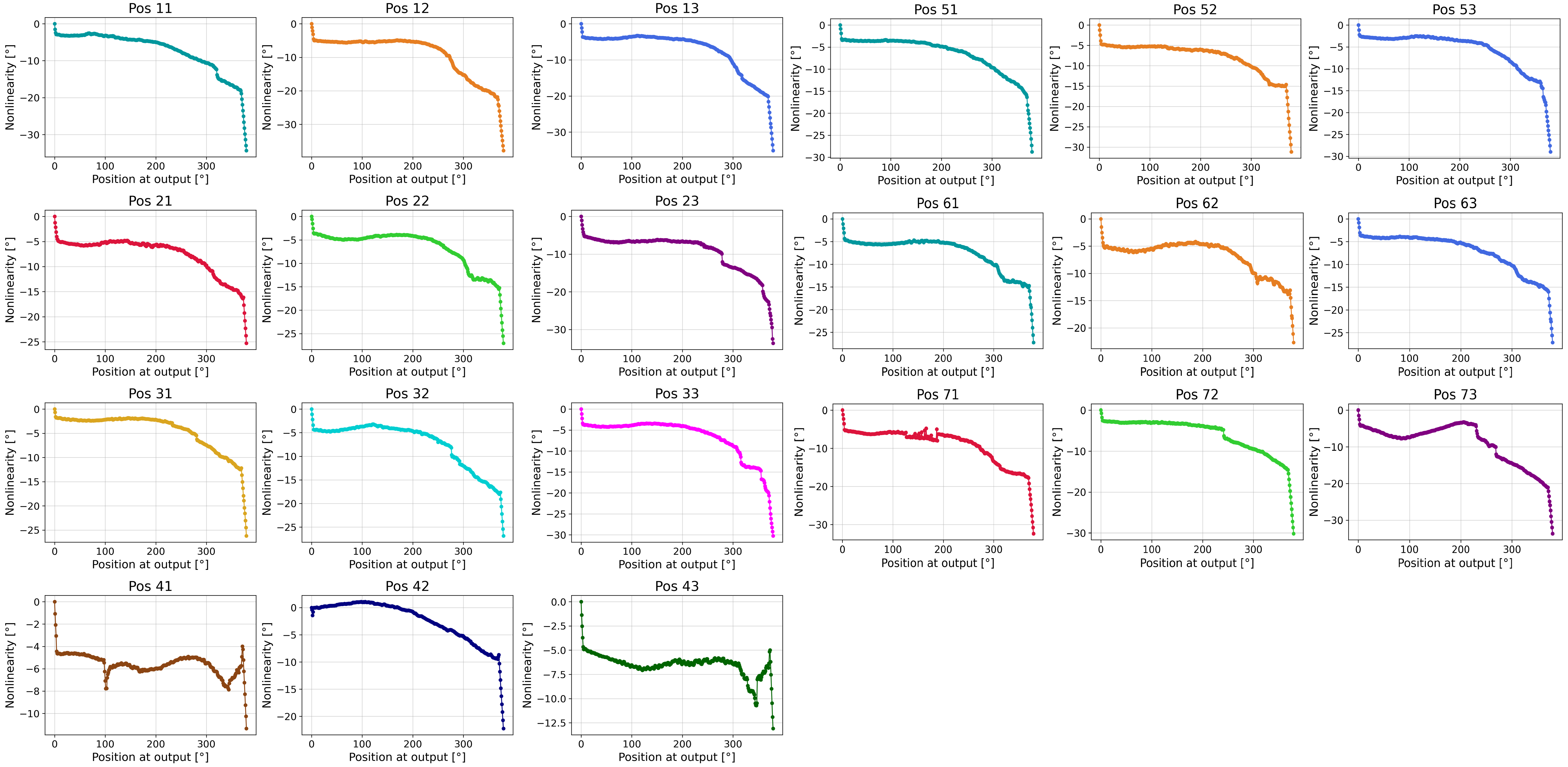}
    \caption{Non-linearity graphs for the alpha arm of each fiber-positioner.}
    \label{fig:nlalpha}
\end{figure}

\begin{figure}[H]
    \centering
    \includegraphics[width=1\linewidth]{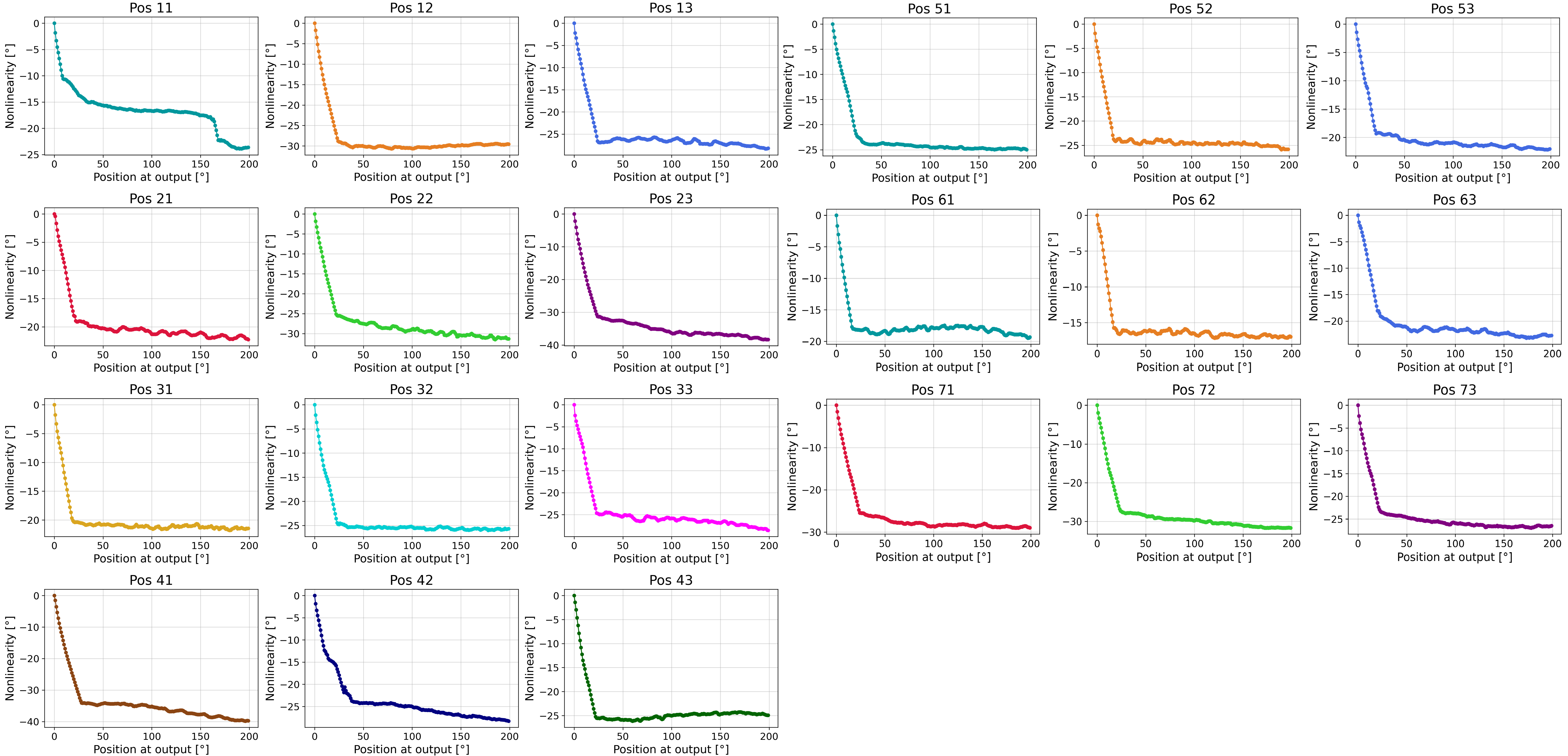}
    \caption{Non-linearity graphs for the beta arm of each fiber-positioner.}
    \label{fig:nlbeta}
\end{figure}

Figures \ref{fig:nlalpha} shows the results obtained for the alpha arm and Figure \ref{fig:nlbeta} for the beta arm. The overall results seem quite promising but further analysis of how to compensate this behavior will be studied. The pronounced slopes on the edges of the plots are attributed to the positioner overcoming the backlash at the hardstop. This behavior is mostly observed in the first 25° to 30° of the output position of the beta arms of the positioners. 

\subsubsection{ARM LENGTH}

\begin{figure}[H]
  \makebox[\textwidth][c]{%
    \begin{minipage}{1\textwidth}
      \centering
      \captionsetup[subfigure]{justification=centering}
      \begin{subfigure}[b]{0.495\textwidth}
        \centering
        \includegraphics[width=\linewidth]{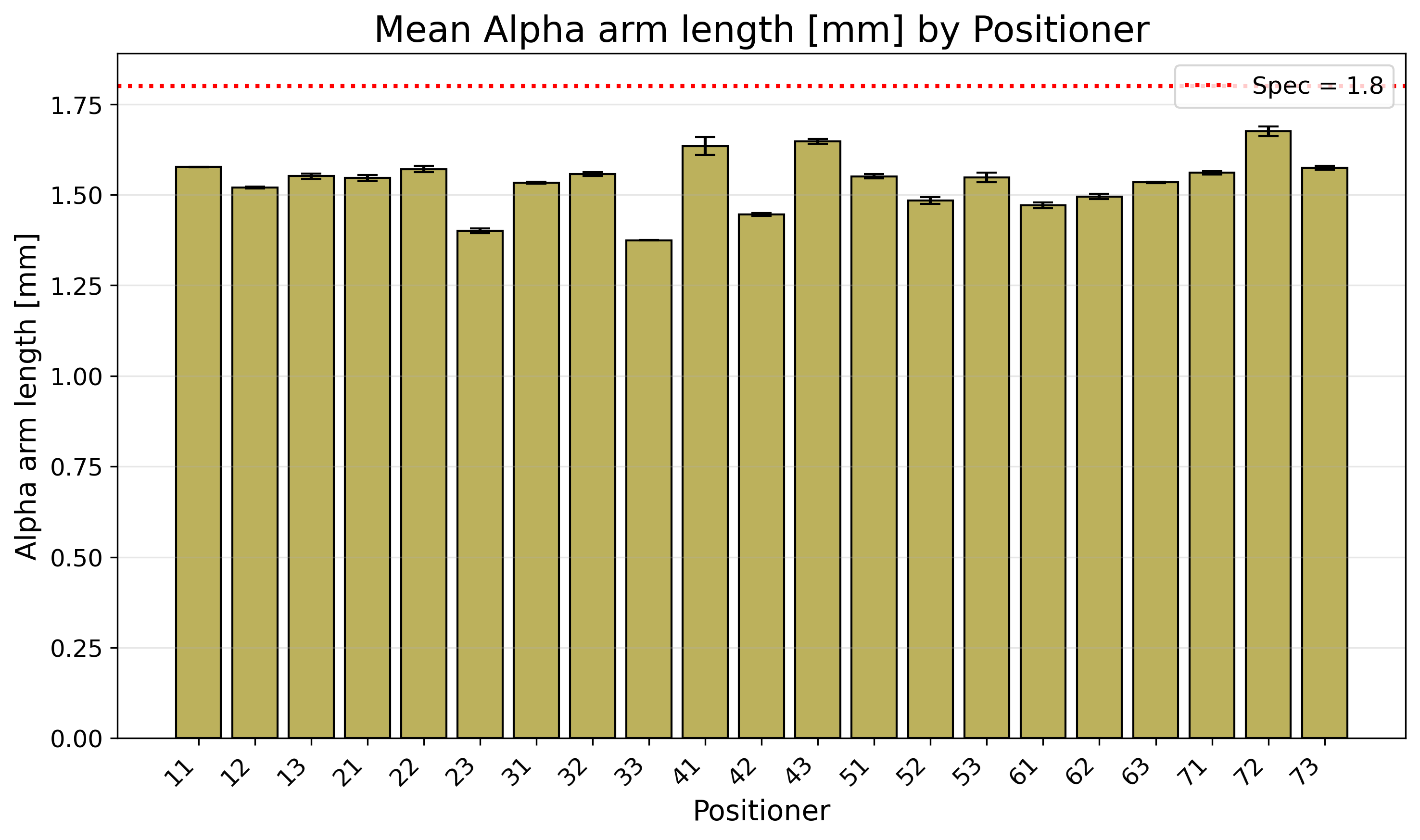}
        \caption{Alpha arm}
        \label{fig:armlengthalpha}
      \end{subfigure}
      \begin{subfigure}[b]{0.495\textwidth}
        \centering
        \includegraphics[width=\linewidth]{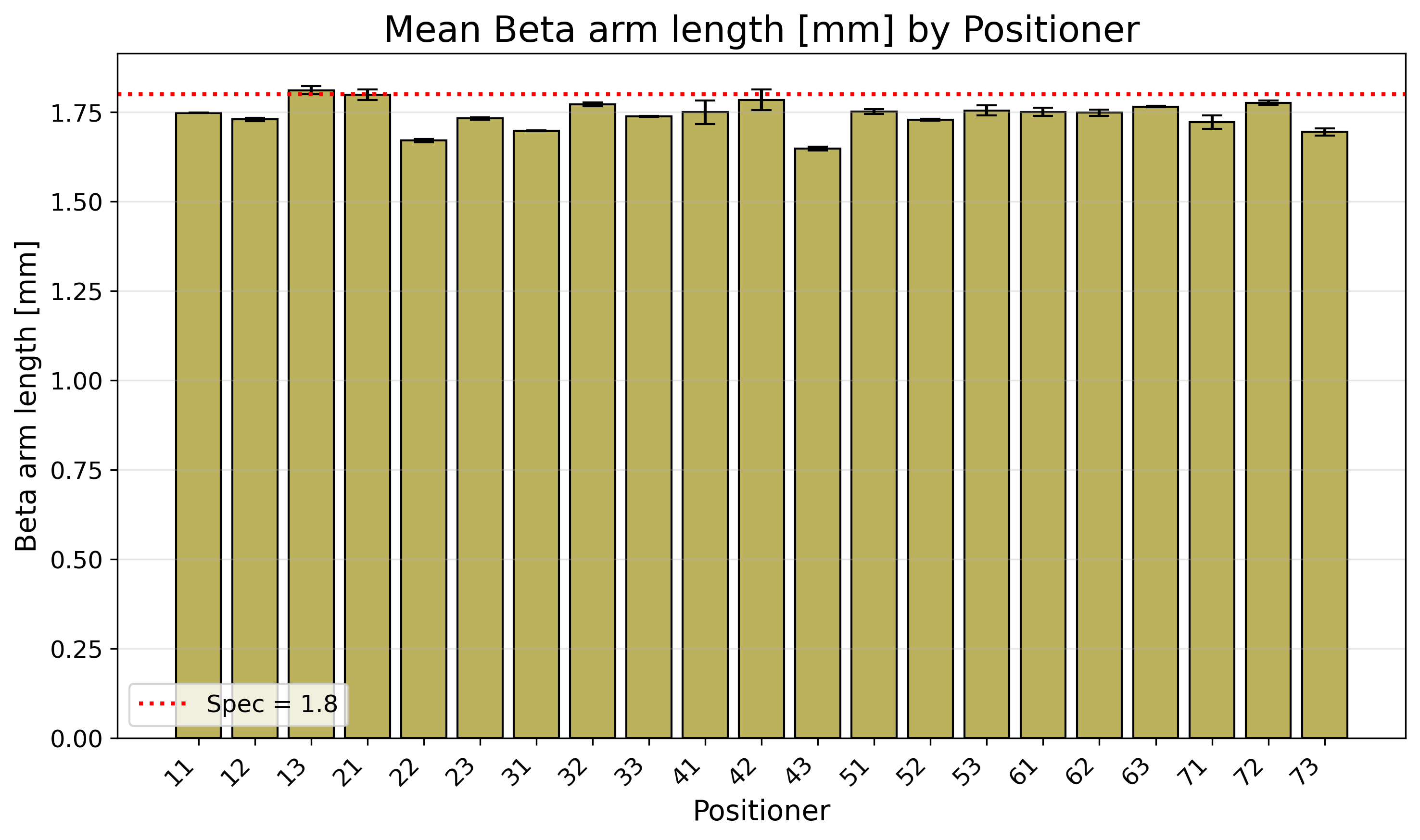}
        \caption{Beta arm}
        \label{fig:armlengthbeta}
      \end{subfigure}
    \end{minipage}
  }
  \vspace{.1cm}
  \caption{Calculated arm length per fiber-positioner. Note: The red dotted line depicts the expected arm lengths of 1.8 mm for each arm.}
  \label{fig:armlength}
\end{figure}

The arm length is determined by tracking the centroids obtained from the motion of each positioner arm independently across its full range of movement. The length of the beta arm is calculated directly as the radius of a circle fitted to the beta arc data points. In contrast, the alpha arm length is derived by fitting beta arcs at multiple alpha positions, extracting the centroids of these fitted circles, and then fitting a circle to those centroids. Figure \ref{fig:armlength} shows the obtained results.

\begin{figure}[H]
    \centering
    \includegraphics[width=0.7\linewidth]{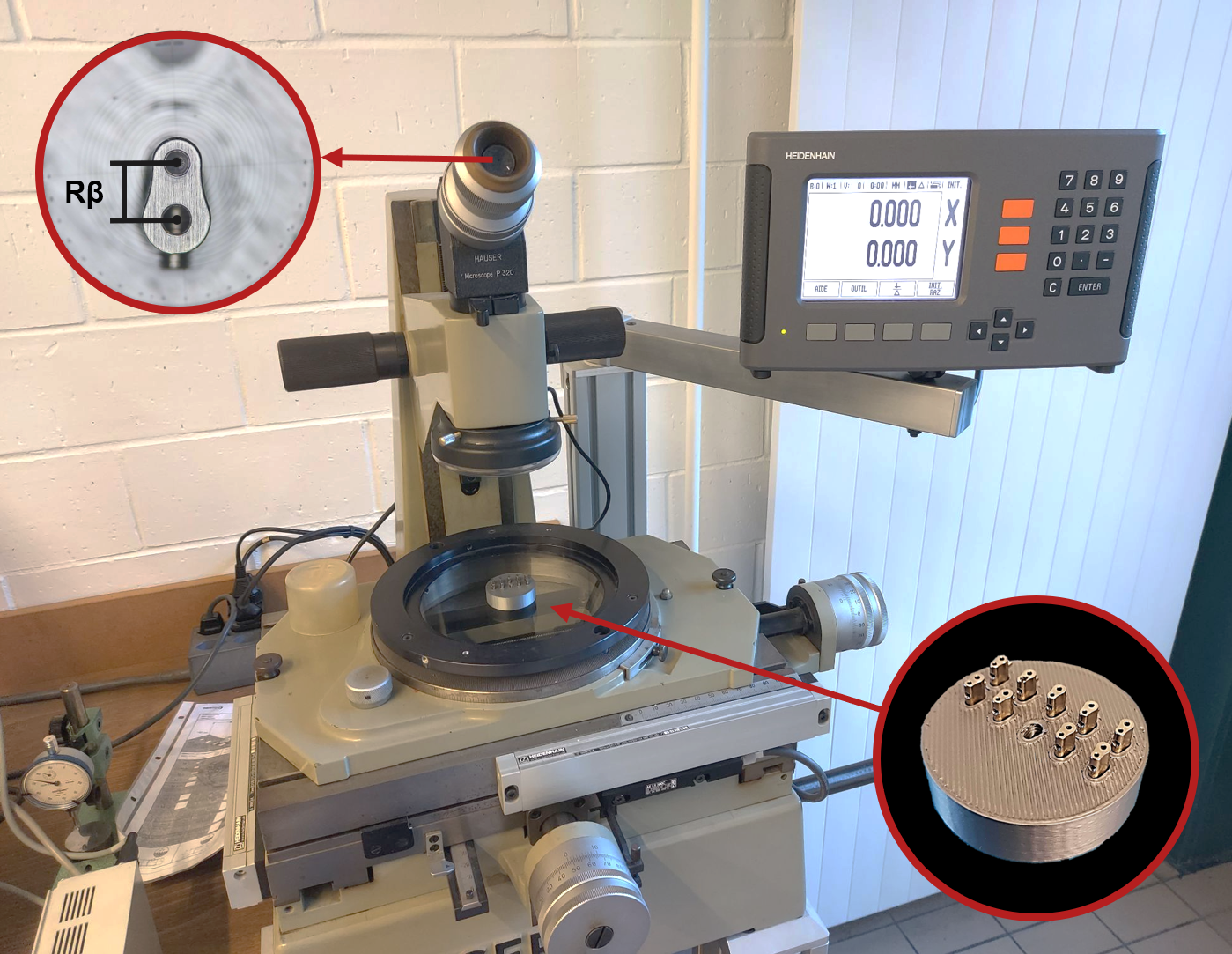}
    \vspace{.1cm}
    \caption{Microscope measurement of additional beta arms from supplier}
    \label{fig:measbetaarms}
\end{figure}

To verify the experimental results obtained for the arm lengths, additional measurements were conducted on a sample of 20 beta arms provided by Orbray. These measurements were carried out using a Hauser P320 microscope, which offers a measurement precision of ±0.015 mm.The measured beta arm distance (R$\beta$) and the measurement setup are shown in Figure \ref{fig:measbetaarms}.

The sample set was divided into two groups based on ferrule hole diameter: 10 arms featured 0.5 mm ferrule holes, while the remaining 10 arms had 0.7 mm ferrule holes. This distribution allowed for a comparative assessment of potential variations associated with different ferrule sizes.

\begin{table}[H]
    \centering
    \caption{Results from Microscopic Measurement of Additional Beta Arms}
    \vspace{1em}
    \label{tab:measbetaarms}
    \begin{tabularx}{0.6\textwidth}{lYYYY}
        \toprule
        \textbf{Arm} & & \textbf{Beta} & \textbf{Beta} \\
        \midrule 
        \textit{Ferrule Hole Diameter}& [mm] &\textit{0,5} & \textit{0,7}\\
        \midrule \midrule
        Mean Arm Length& [mm]&1,84 & 1,82 \\
        Min Arm Length& [mm]&1,80 & 1,80\\
        Max Arm Length& [mm]&1,86 & 1,87\\
        \bottomrule
    \end{tabularx}
\end{table}

The obtained results are presented in Table \ref{tab:measbetaarms}. Further testing needs to be done to evaluate physical measurements of the alpha arms.

\subsubsection{MOTION RANGE}

\begin{figure}[H]
    \centering
    \includegraphics[width=1\linewidth]{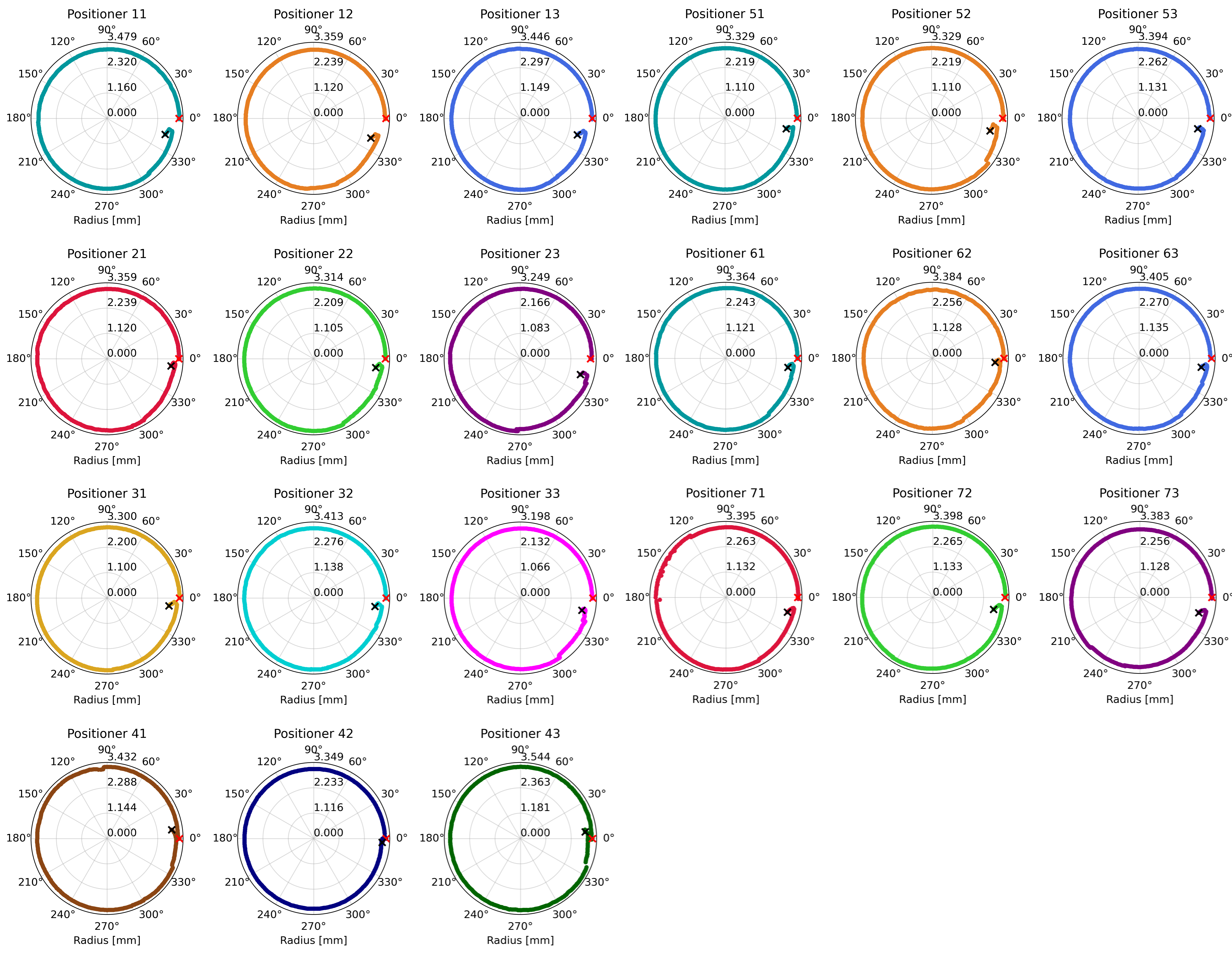}
    \caption{Alpha arm motion range per fiber-positioner. Note: Red 'x' is the start, and black 'x' is the end. Note: Desired range = 370 degrees.}
    \label{fig:motionrange}
\end{figure}

\begin{figure}[H]
    \centering
    \includegraphics[width=1\linewidth]{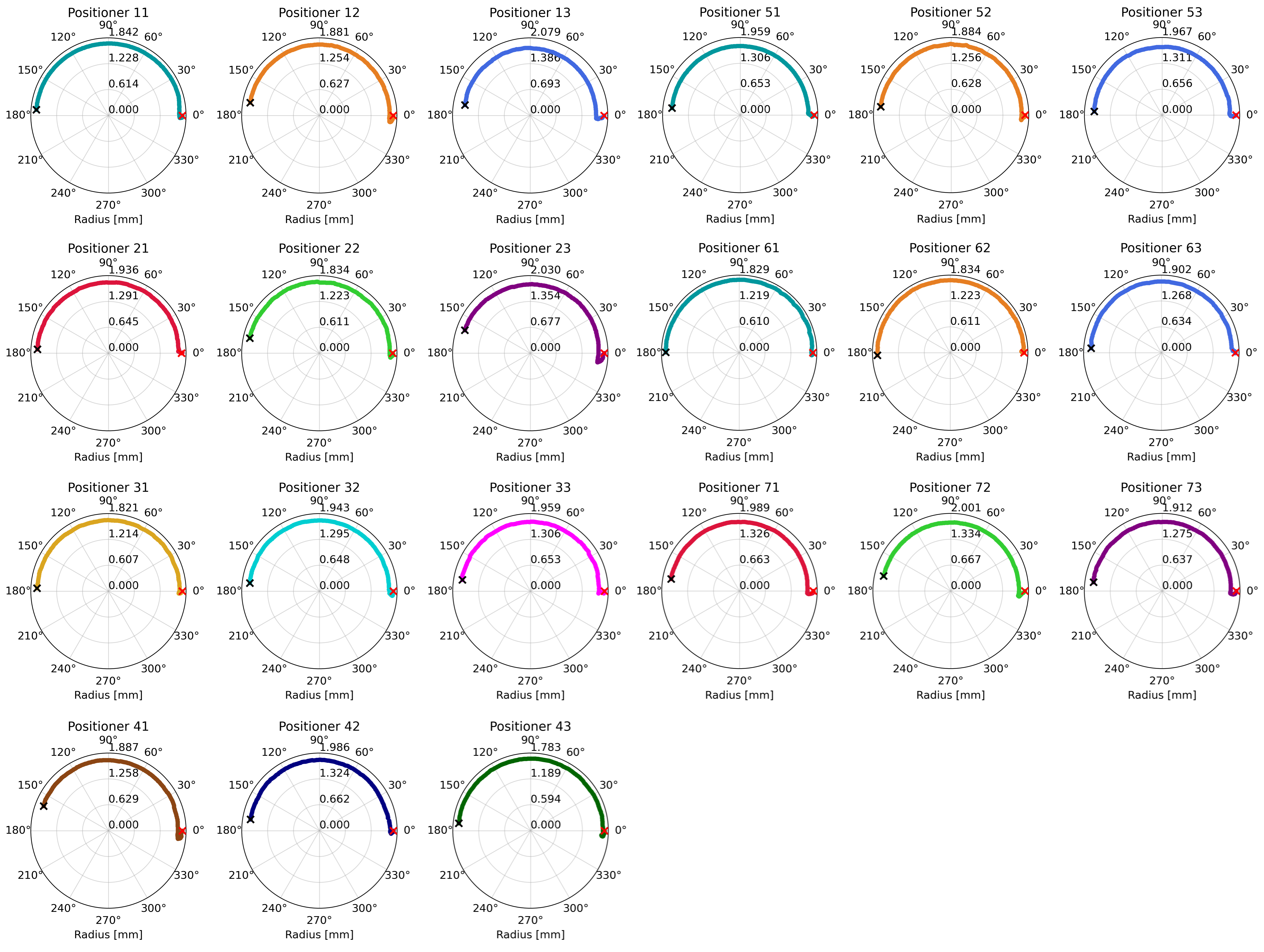}
    \caption{Beta arm motion range per fiber-positioner. Note: Red 'x' is the start, and black 'x' is the end. Note: Desired range = 190 degrees.}
    \label{fig:motionrange2}
\end{figure}

The motion range of each fiber-positioner was tested by recording the movement from each one of its arms from its hard stop in one direction to its hard stop in the opposite direction. Figures \ref{fig:motionrange} and \ref{fig:motionrange2} shows the results obtained for the alpha and beta arms respectively.

\newpage
\vspace*{\fill}
\vspace*{\fill}

\newpage
\vspace*{\fill}
\vspace*{\fill}
\newpage

\begin{table}[H]
    \centering
    \caption{Motion range summary and comparison between measured values and the desired performance.}
    \vspace{1em}
    \label{tab:motionrange}
    \begin{tabularx}{0.55\textwidth}{p{0.7cm} Y Y Y Y}
    \toprule
        \textbf{Pos} & \textbf{Measured alpha motion range [°]} & \textbf{Measured beta motion range [°]}  \\ \midrule
        11 & 344.67 & 175.38 \\
        12 & 341.18 & 169.45 \\
        13 & 343.88 & 170.80 \\
        21 & 353.72 & 176.76 \\
        22 & 351.96 & 167.71 \\
        23 & 345.35 & 160.67 \\
        31 & 352.80 & 177.56 \\
        32 & 352.14 & 173.33 \\
        33 & 348.75 & 170.45 \\
        41 & 367.65 & 159.23 \\
        42 & 356.77 & 170.71 \\
        43 & 365.91 & 174.06 \\ 
        51 & 350.24 & 174.02 \\
        52 & 347.85 & 173.16 \\ 
        53 & 350.03 & 176.95 \\ 
        61 & 351.74 & 179.68 \\ 
        62 & 356.32 & 182.04 \\ 
        63 & 351.85 & 176.27 \\ 
        71 & 346.60 & 170.01 \\
        72 & 348.82 & 167.33 \\ 
        73 & 345.29 & 172.56 \\ \bottomrule
    \end{tabularx}
\end{table}

Table \ref{tab:motionrange} showcases a summary of the results obtained from the previous plots. Overall the results and the behavior observed in the figures show that some of the positioners present difficulties in achieving their motion range. In addition, noticeable radial shifts can be observed in most positioners which might impact their positioning performance.

\subsection{ANGULAR TILT}

\begin{table}[H]
    \centering
    \caption{Tilt testing results per fiber-positioner. Note: The deviation column is calculated as the difference between the Residual Sum of Squares (RSS) of both arms and the desired performance \(\le \) 0.4°}
    \label{tab:tilttest}
    \vspace{1em}
    \begin{tabularx}{0.95\textwidth}{p{0.7cm} Y Y Y Y Y}
    \toprule
        \textbf{Pos} & \textbf{Tilt $\theta$ [°]} & \textbf{Tilt $\phi$ [°]} & \textbf{Direct Sum [°]} & \textbf{RSS [°]} & \textbf{Deviation [°]} \\ \midrule
        11 & 0.32 & 0.27 & 0.59 & 0.42 & 0.02 \\ 
        12 & 0.34 & 0.27 & 0.61 & 0.43 & 0.03 \\
        13 & 0.25 & 0.28 & 0.53 & 0.37 & -0.03 \\ 
        21 & 0.40 & 0.22 & 0.62 & 0.46 & 0.06 \\ 
        22 & 0.47 & 0.27 & 0.74 & 0.54 & 0.14 \\
        23 & 0.21 & 0.36 & 0.57 & 0.41 & 0.01 \\ 
        31 & 0.35 & 0.26 & 0.61 & 0.44 & 0.04 \\ 
        32 & 0.44 & 0.21 & 0.65 & 0.49 & 0.09 \\ 
        41 & 0.06 & 0.14 & 0.20 & 0.15 & -0.25 \\
        42 & 0.46 & 0.40 & 0.86 & 0.60 & 0.20 \\ 
        51 & 0.41 & 0.25 & 0.66 & 0.48 & 0.08 \\
        52 & 0.51 & 0.36 & 0.87 & 0.62 & 0.22 \\ 
        53 & 0.44 & 0.23 & 0.67 & 0.50 & 0.10 \\ 
        61 & 0.56 & 0.36 & 0.92 & 0.66 & 0.26 \\ 
        62 & 0.15 & 0.39 & 0.54 & 0.41 & 0.01 \\
        63 & 0.37 & 0.33 & 0.70 & 0.50 & 0.10 \\ 
        71 & 0.37 & 0.27 & 0.64 & 0.45 & 0.05 \\ 
        72 & 0.20 & 0.20 & 0.40 & 0.28 & -0.12 \\ 
        73 & 0.53 & 0.18 & 0.71 & 0.56 & 0.16 \\ \bottomrule
    \end{tabularx}
\end{table}

It can be observed in these results that reaching the desired 0.4° is quite challenging as most of the positioners presented values above this threshold.

\subsection{SUMMARY}

\begin{table}[H]
    \centering
    \caption{Comparison matrix showing the results of the Orbray prototype and the desired performance for Stage-5 instruments\cite{SilberModule}. Results are shown for the best, worst and the average of the performing fiber-positioner units in the module. Note: Combined values for Repeatability and Datum Repeatability are calculated as the quadrature sum of the $\alpha$ and $\beta$ components.}
    \vspace{1em}
    \begin{tabularx}{\textwidth}{p{5.5cm} Y Y Y Y}
        \toprule
        \textbf{Parameter} & 
        \textbf{Desired Perf.} & 
        \textbf{Best Unit} & 
        \textbf{Worst Unit} &
        \textbf{Average}\\
        \midrule\midrule
        
        Repeatability $\alpha$ ($\mu$m RMS) & $\leq$6 & 1.44 & 77.87 & 18.23\\
        Repeatability $\beta$ ($\mu$m RMS) & $\leq$6 & 0.81 & 7.61 & 2.41 \\
        Repeatability (Comb.) ($\mu$m RMS) & -- & 1.65  & 78.24 & 18.39\\
        
        \midrule
        
        Datum Repeatability $\alpha$ ($\mu$m RMS)  & $\leq$50 & 16.89 & 63.24 & 38.48\\
        Datum Repeatability $\beta$ ($\mu$m RMS)   & $\leq$50 & 1.10 & 31.99 & 9.00\\
        Datum Rep. (Comb.) ($\mu$m RMS)   & -- & 16.93 & 70.87 & 39.52\\
        
        \midrule
        
        Backlash $\alpha$ (deg) & $\leq$5 & 2.48 & 7.64 & 3.30\\
        Backlash $\beta$ (deg) & $\leq$5 & 5.60 & 15.57 & 11.32\\
        
        \midrule
        
        Angular Tilt (deg) & $\leq$0.4 & 0.15 & 0.66 & 0.46\\
        
        \bottomrule
    \end{tabularx}
    \label{table: comparison}
\end{table}

A summary of the performance from this prototype is shown in Table \ref{table: comparison}. The results are compared with the desired performance for Stage-5 telescope projects. These results give us an insight of the metrics that must be improved in further iterations to achieve the desired performance.

\section{DISCUSSION}

An improvement in the results of the positioners when compared to the previous prototype iteration from Ref \citenum{Galal2025} was observed. Figure \ref{fig:repalpha} shows that the XY repeatability of most positioners in the module is below the desired performance threshold. However, among the positioners that exceed this threshold for the alpha arm, five stand out because their repeatability values are exceptionally high and show large variance. This may indicate mechanical problems in the alpha arm actuation of those units. The results obtained from these positioners highly impact the average from the module (Table \ref{table: comparison}). Without considering these positioners the average combined repeatability for the alpha arm would be 5.65 $\mu$m which is below the threshold. In the case of the beta arm results from Figure \ref{fig:repbeta}, only a couple of the positioners had values that exceeded the threshold. Although their values are above the specified limit, they do not reach the levels observed from those in the alpha arm results. 

Figure \ref{fig:datumalpha} shows the datum repeatability results for the alpha arm of each positioner. As observed, most of the positioners are within the desired performance for the alpha arm with only a few anomalies that may improve in the future as further tests are performed. An important issue currently being addressed with the manufacturer is the malfunction of the alpha hardstop in Positioner 33. This malfunction allowed the positioner to move without a physical reference, which subsequently damaged both the positioner and the attached optical fiber, preventing further testing. As a result, tilt-test results for this positioner are not available. This behavior will be studied further to make improvements that prevent this to happen in next iterations. In contrast, the beta arm results from Figure \ref{fig:datumbeta} show that all positioners accomplish the desired performance being only Positioner 41 the one that presents a high variability when compared to the rest.

The backlash results for the alpha arm shown in Figure \ref{fig:backlashalpha} show that most positioners are well within the requirement with the exception of Positioner 23, whereas the beta backlash results from Figure \ref{fig:backlashbeta} are significantly higher in all positioners. The backlash can potentially be calibrated within the software. Given the gearstack and the complex mechanics of the trillium design, it might be difficult to attain better performances, and the solution would be adapting the software rather than changing the hardware. The degree on which the impact of this high backlash in the beta arm can be compensated via software solutions will be studied in the future as its an important aspect for the accuracy of these systems.

These observations required further investigation. Therefore, a more detailed examination of the positioners during operation was carried out. During commanded motion, it was observed that some positioners exhibited non-smooth behavior as shown in the motion range profiles from Figure \ref{fig:motionrange} and Figure \ref{fig:motionrange2} which might impact the performance. It should be noted that most of the positioners do not fully meet the desired motion range, although their achieved ranges are close to the required values. Even though, these radial shifts along their motion range affect positioning performance and can be attributed to both, manufacturing and design aspects that need to be improved further. 

The lengths of the alpha and beta arms are calculated from an alpha circle created by fitting the centroids of beta arcs along the motion range of the positioner. Figure \ref{fig:armlengthalpha} shows the values calculated for the alpha arm lengths, and it seems that the arm lengths are quite different from what the expected value (1.8 mm for each arm). This result could be due to the arms actually being shorter than what is expected or it could be resulting from the non-smooth motion of the robots which is leading to an erroneous measurement. This needs to be investigated further and discussed with the manufacturer. In the case of the beta arms (Figure \ref{fig:armlengthbeta}) while the measured values seem closer to the expected values there are still noticeable differences. The microscope measurements of additional beta-arm samples, summarized in Table \ref{tab:measbetaarms}, provide insight into the impact of the previously mentioned mechanical issues on the motion of the positioner and, consequently, on the measured arm lengths. These results suggest that the observed discrepancies are more likely associated with issues related to the motion of the positioners than with deviations in the manufactured beta arms.

Figures \ref{fig:nlalpha} and \ref{fig:nlbeta} show the non-linearity plots for the different positioners. The alpha arm profiles exhibit a common oscillatory trend, suggesting a systematic behavior across the tested units. Since non-linearity should be below 0.8°, the results need to be analyzed further. As the non-linearity can be calibrated if it exhibits a repeatable pattern it is necessary to implement and evaluate this compensation to assess the degree of improvement it can bring to the positioner's performance during operation. 

In addition to the previously mentioned damage to Positioner 33, Positioner 43 exhibited a mechanical issue that prevented continued operation of its beta arm, leaving it in a partially extended position. For this reason, these two positioners were excluded from the angular tilt test and will not be included in subsequent tests. Nevertheless, both units will be further examined to determine the root cause of the failures and to propose design or manufacturing modifications for the next prototype, with the aim of reducing or eliminating the likelihood of these problems.

Based on the RSS values of the angular tilt results reported in Table \ref{tab:tilttest}, only 3 of the 19 tested positioners achieved a tilt below the desired threshold of 0.4 deg. The RSS value is used for the comparison because the direct sum represents a worst-case scenario where tilt from both axes is aligned. Since the axes are not perfectly aligned, RSS appears more representative. While some units showed values only slightly above this limit, more than one third of the positioners exhibited tilt values at least 0.1 deg above the specified threshold. These results, together with the observed flexibility of the positioner's mechanical structure, indicate that besides improvements in the mechanics, reinforcements or material modifications might be required to increase the stiffness of the positioners.

\section{CONCLUSION}

Overall, this prototype showed encouraging results, with some positioners reaching acceptable performance levels. At the same time, the behavior of other units provided useful insight into the aspects of the design that need further refinement. A small number of positioners showed more concerning issues, which are currently being addressed in collaboration with the manufacturer, Orbray. Additional testing such as accuracy, thermal, lifetime and orientation\cite{wuthrich2026gravity} tests will be performed to better understand the behavior of these systems and to evaluate the proposed improvements. For the next prototype, mechanical improvements will be especially important to improve beta arm backlash, alpha arm repeatability, and angular tilt.

\acknowledgments 
 
The authors would like to acknowledge the Swiss National Science Foundation (SNSF) for supporting this work through the Funding LArge international REsearch projects (FLARE) grant. Additionally, the authors also gratefully acknowledge LBNL for the design of the mechanism, Orbray for manufacturing the prototype and Dr. Luzius Kronig for the many valuable discussions and his help throughout this work. Finally, the authors warmly acknowledge the EPFL workshops for their expertise and care in producing the precision components used in the test benches.

\bibliography{report} 
\bibliographystyle{spiebib} 

\end{document}